# Micro-plasticity and recent insights from intermittent and small-scale plasticity

R. Maass[1] and P.M. Derlet[2]

[1]Department of Materials Science and Engineering, University of Illinois at Urbana-Champaign, Urbana, IL 61801, USA
[2]Condensed Matter Theory Group, Paul Scherrer Institute, 5232 Villigen, Switzerland

Prior to macroscopic yielding, most materials undergo a regime of plastic activity that cannot be resolved in conventional bulk deformation experiments. In this pre-yield, or micro-plastic regime, it is the initial three dimensional defect network that is probed and the intermittently evolving microstructure admits small increments in plastic strain. By reducing the sample size, this intermittent activity becomes increasingly apparent and can be routinely observed through small-scale mechanical testing. In some cases, the intermittent activity was shown to exhibit aspects of scale-free behavior, prompting a paradigm shift away from traditional microstructure-dependent unit mechanisms that may be associated with a well defined length and stress scale. In this article, we discuss and review connections between classical micro-plasticity and intermittent flow across all length scales, with the aim of highlighting the value of miniaturized testing as a means to unravel this very early regime of bulk plasticity.

a) electronic mail: rmaass@illinois.edu, peter.derlet@psi.ch

## 1 Introduction

Mechanical straining of a macroscopic metallic material results in some elastic deformation, followed by either abrupt brittle-like failure, or by a gradual transition from macroscopically elastic to macroscopically plastic deformation via a yield transition. This general behavior prevails irrespective of microstructure, encompassing single crystals, polycrystals, nanocrystalline metals, high-entropy alloys and even metallic glasses. Historically and somewhat arbitrarily, the transition from macroscopic elasticity to yield is defined as the point along a deformation curve at which a plastic strain of 0.2% is achieved [1]. Depending on the density of the initial defect structure and the mobility of the fundamental carriers of plastic flow, there is always some small (often unresolvable) amount of permanent deformation throughout the macroscopically elastic regime. This is even the case for almost instantaneous macroscopic failure at the yield transition, which can be observed for many high-entropy alloys or metallic glasses. The other extreme are high-strength nanocrystalline metals, where deviations from perfect macroscopic-linear behavior are detected at very low stresses [2], evidencing plastic activity at fractions of the macroscopic yield or flow strength. In both cases, it is the deviations from the perfect linear-elastic initial loading behavior that often cannot be resolved with conventional macroscopic testing, but which are present in all metallic materials containing mobile defect populations. These small deviations from Hooke's law, which can occur at a fraction of the macroscopic elastic limit, give rich insight into the initial defect structure, the operative microscopic plastic mechanisms, intermittent plastic flow, and how a defect ensemble evolves towards the yield transition.

Plasticity prior to macroscopic yield is often referred to as the micro-plastic or micro-strain regime; a regime of deformation that experienced intense research efforts in the 1960s and early 1970s, using instrumentation which could achieve displacement resolutions approaching magnitudes of only a few Burger's vectors. With the advent of modern small-scale mechanical testing and characterization, displacement deviations away from perfect elastic behavior are now routinely detected with a similar resolution. Such discrete, non-linear, behavior is however often treated as a large stress-strain evolution and therefore within the framework of a conventional bulk stress-strain curve. It is, however, these small displacement increments in small-scale deformation experiments that define micro-plasticity in bulk crystals: local plastic transitions that individually do not lead to statistically meaningful changes in dislocation structure. A natural question is consequently whether there are any fundamental similarities in the deformation physics of bulk micro-plasticity and small-scale plasticity, or is the only apparent connection a required similar displacement resolution to resolve plastic activity?

The purpose of this article is to investigate this question and to establish a connection between the traditional concept of micro-plasticity and the contemporary insights gained from intermittent and small-scale plasticity. In doing this, a contrasting view emerges – classical (micro-)plasticity is based on the understanding of finite length scales, unit plastic mechanisms, and mean properties of dislocation structures, whereas many contemporary results from bulk and small-scale plastic straining of both crystals and metallic glasses indicate intermittent plasticity that can be well described by scale-free distributions (pure power-law or truncated power-law) that per definition do not have a defined mean.

In the remainder of this section we define how we view micro-plasticity and why it is interesting within a modern context. In Section 2, some early observations of bulk crystalline micro-plasticity under different loading modes will be reviewed, discussing how micro-plasticity was traditionally understood in terms of local mechanisms and well-defined average quantities. Section 3 introduces micro-plasticity in metallic glasses. Subsequently, in Section 4, we address intermittent plastic flow in bulk and small-scale systems, and how these insights shift focus from well-defined mean quantities to scale-free aspects of plasticity. It is at this point that we ask and discuss the question to what extent the discrete plastic activity seen at the small scale is a manifestation of the micro-plastic regime prior to yield in a bulk material. Future research directions and open questions are presented in Section 5.

*1.1 What is micro-plasticity?*

Deduced directly from a macroscopic stress-strain curve, micro-plasticity can be simply defined as all inelastic deformation activity occurring during the pre-yield region, also called the micro-strain regime. Such deviations from elasticity in the pre-yield regime can, for example, be seen in time-dependent loading, open-loop loading hysteresis, or as distinctly visible displacement jumps. This perspective does not include anelastic reversible (closed-loop loading hysteresis) behavior, as was frequently observed during micro-plasticity studies of both fcc and bcc metals. As such, the observation of micro-plasticity is a displacement (strain) resolution problem that was already discussed in the 20s-30s by numerous authors [3-6]. In fact, even then, it was questioned if yield was at all free from the measurement system used to detect it [4]. One way to circumvent the strain resolution problem in macroscopic deformation, is to directly record optical changes in a transparent crystal [7], to listen to emitted sound by ear [8], or to record elastic waves via acoustic emission (AE) sensing, where local plastic transitions that do not give rise to measurable deviations from linear elasticity are seen as acoustic pulses. Several examples are found for both crystalline [9-11] and amorphous metals [12-14], where the appearance of crackling noise or acoustic emission pulses in the pre-yield region clearly indicate operative plastic processes.

This observation leads to the second possible definition of micro-plasticity, which entirely relies on a microstructural picture. According to this viewpoint, micro-plasticity encompasses plastic deformation of a crystal without significant alternation of the initial microstructure or the emergence of new internal length scales. An example of this is the observation that AE pulses vanish up to the previous highest stress level during load-unload cycles in the macroscopically elastic regime, indicating the exhausting of local plastic process up to the stress of the foregoing loading [12, 15]. Interestingly, this pre-yield micro-plastic activity does not change the yield strength noticeably and it is also said to not alter the internal length scales of the pre-existing microstructure [16-18]. Being less direct and more difficult to verify than a definition based on deviations from elasticity, this microstructural definition relies on knowing, and being able to trace, the properties of a pre-existing defect structure during plastic evolution. For crystalline metals, parameters such as dislocation density, mean dislocation-boundary spacing, and other internal length scales are thus the relevant measures.

A deviation from a perfect elastic response at very low stress, as recorded with strain measurements, includes smooth plastic creep strain and plastic uniaxial strain, as well as intermittent plastic strain jumps. In particular the numerous early uniaxial micro-strain experiments gave insights into pre-yield phenomenon and therefore micro-plasticity. Figure 1 displays schematically a stress-strain curve (i) and the stress-strain response obtained during classical micro-yield experiments (ii-iv), highlighting the regime of micro-plasticity between $\tau_{\mu p}$ and the yield stress $\tau_Y$.

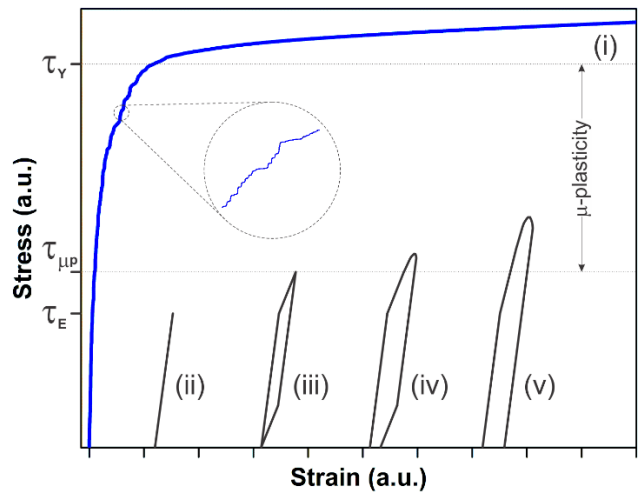

Figure 1: Schematic stress-strain curve (i) and different responses obtained during pre-yield testing: elastic limit (ii), anelastic hysteresis (iii), permanent micro-plastic deformation (iv), half-loop response during micro-plastic cyclic loading (v). The zoom-in between $\tau_{\mu p}$ and $\tau_Y$ highlights intermittent micro-plastic deformation.

Independent of the pre-existing defect structure, a true elastic limit, $\tau_E$, can be defined (ii). Whilst this limit clearly depends on the experimental strain resolution, typical values for the so-called micro-yield stresses obtained from high resolution (~nm) extensometry, etch-pit studies, or x-ray topography on bulk single crystals are ~2 MPa (Mo, W, Fe, [19]), ~1-4 MPa (Cu, [16, 20]), ~0.2-0.5 MPa (Zn, [21, 22]), ~ 0.5 MPa (Be, [23]), and ~1.5 MPa (Cu, [24]), ~2 MPa (Al, [25]), ~2-30 MPa (Fe, [19, 25, 26]) was found for polycrystals. These values are thus distinctly lower than corresponding levels of a macroscopic $\tau_Y$ (~10-30 MPa (Cu [16, 24]), ~80-165 MPa (Fe [19, 26]), ~10 MPa (Be, [23]), ~500 MPa (W, [19])). A first manifestation of deviations away from a linear elastic response is the formation of a closed loop upon unloading (anelastic response, (iii)) for crystals with a pre-deformation history. The enclosed loop area is dependent on the initial dislocation density, temperature, strain amplitude, and it has been concluded that the dislocation structure remains identical during cyclic loading as long as the loop remains closed, that is below $\tau_{\mu p}$ [27, 28]. Despite not being a central aspect of this article, the fully recoverable nonlinear behavior is an interesting observation, since it probes the frictional lattice stress from the energy loss via [22, 29]: $W = \oint(\tau_G + \tau_F)d\gamma$, where $\tau_G$ is the stress opposing dislocation motion due to elastic interaction, $\tau_F$ is the friction stress, and $\gamma$ is the plastic shear strain. Using this approach, one considers short range movement of dislocations by thermally activated processes that cause the energy loss, $W$. The limit of the closed hysteresis loop marks the transition from reversible dislocation-segment bowing to irreversible dislocation motion and thus the onset of the micro-plastic regime (iv). Passing $\tau_{\mu p}$, permanent plastic strain can be recorded. In a graph summarizing $W$ as a function of $\gamma$ ($\gamma$ is here the shear strain deviation from linear elasticity for closed loops, and the non-recoverable strain for open loops), the onset of the micro-plastic regime will be indicated by a deviation from linearity, and the slope prior to that point yields the friction stress which is typically of the order of some kPa [22] to MPa [19, 23]. The micro-plastic regime itself has gained considerably less attention than the micro-strain regime below $\tau_{\mu p}$. Above this stress, closure of the load-unload loop is absent, with selected evidence for time-dependent (partial) loop closure [22]. This evidences dislocation creep away from their obstacles due to the interacting stress fields. Thus, after a sufficient amount of waiting time, there is a non-linear strain recovery similar to closed loop formation during quasi-static loading, but also a resulting permanent plastic micro-strain.

In addition to the large body of micro-strain investigations, micro-plastic permanent discrete deformation has also been resolved under many different loading modes. As early as in the 1930s numerous experiments hinted that dislocation plasticity is an intermittent process leading to jerky flow [30-32]. This is not unique to macro-plastic flow, but with sufficient high strain resolution, clear sudden jumps of displacement can be observed at stresses well below the macroscopic yield [20, 21, 33], as schematically depicted in the blow-up of the schematic stress-strain curve in Figure 1.

Finally, Figure 1 also displays half a stress-strain loop from an internal friction measurement (v), which is shown for completeness, as it essentially is the high frequency version of the previously discussed closed loop response during quasi-static uniaxial micro-strain testing.

*1.2 What makes micro-plasticity interesting?*

In the following sections it will be demonstrated that the study of the micro-plastic regime has a long history, the consequences of which have influenced and given birth to a fundamental understanding of the microscopic origins of plasticity. Indeed, the discussion around the early experiments of Orowan and Becker [31], Chalmers [4], and Tinder [21], which could probe micro-plasticity using remarkably high strain resolutions, forms the basis of our concepts of unit-scale mechanisms, thermal activation, dislocation organization and the (instrumentation dependent) transition and measurement of yield. To see this, it is worth recalling some notable text in these papers – work which will be reviewed in more detail in the proceeding sections.

In the 1932 work of Orowan and Becker [31], which involved temperature dependent creep tests showing discrete changes in elongation, the stated open questions to be addressed were:

1. "Wie entsteht ein lokales Gleiten und wodurch ist die Zahl der sekündlich angestossenen Gleitvorgaenge gegeben? (How does local gliding begin and what determines the number of glide processes that initiate every second?)"
2. "Wie wächst sich das lokale Gleiten zum elementaren Gleitakt aus und wodurch ist der Ablauf (Gleitgeschwindigkeit und Abgleitungsbetrag) eines elementaren Gleitaktes bestimmt? (How does the local gliding grow into an elementary act of gliding, and what determines the development (rate of gliding and extent of the individual act) of the elementary glide act?)"

Here the English translations are taken from Nabbaro and Argon [34]. In reading these, one notices the early reference to glide – the quantitative understanding of the dislocation network was only then beginning to emerge through the parallel work of Taylor and Polanyi (see for example [35, 36] and references therein) – and the concept of a unit scale dislocation mechanism. Indeed, it was only a short time later that Orowan produced his recognized theory work on the thermal activation of dislocation glide [37].

Turning to the 1936 work of Chalmers which investigated pre-yield plasticity in tin, and lead to the early usage of the term micro-plasticity [4], he writes in the discussion "It is suggested that if a macroscopic stress less than the critical stress be applied, it will still be possible for dislocations to move owing to the thermal agitation producing local increases of stress, but that this will only happen infrequently, as it depends on the thermal movement being sufficient to increase the effect of the applied tension up to the value required to cause movement of the dislocations. Each time this happens, a definite extension takes place, and since the number of remaining dislocations is thus continually decreased, the rate of extension gradually decreases. This explains the experimental fact that the rate of creep depends on the stress, and under constant stress decreases with increasing length, giving a finite limit for the total creep." Like the early work of Orowan and Becker, the strong connection between temperature and intermittency is here emphasized, and also that micro-plasticity can become exhausted suggests a rather discrete relationship between the structure prior to loading and its micro-plastic response.

Subsequently, we proceed with a more modern and lesser known work of torsion experiments reported by Tinder and co-workers [21]. The discrete plasticity they saw, in between perfect regions of elasticity, as a function of whether the sample had been pre-strained or not, naturally lead them to conclude "… an important fraction of the total strain, in the initial stages of deformation, involved motion of a few favorably situated dislocation segments through distances large enough to form new interactions with other elements of the three dimensional network. If this were so, then most elements of the network must have been relatively immobile, making little or no contribution to the strain".

This again underlies that in the micro-plastic regime of deformation, the initial structure of the material has a well-defined meaning as a reference structure within which plasticity occurs – micro-plasticity directly probes the initial three dimensional structure giving insight into its stability, and the unit plastic mechanisms. On the other hand, macro-plastic flow originates from a driven structure – it is emergent – and thus has less of a direct relation to the initial microstructure. Of course, the initial structure determines the ultimate flow properties, but the deformation to get there (to yield and beyond) must also be considered when discussing the underlying plastic mechanism contributing to plastic flow. As will be shown, this aspect of micro-plasticity is central to the methods of mechanical spectroscopy and creep/stress-relaxation tests performed at stress levels well below that of the yield stress.

The second, more contemporary issue which makes micro-plasticity interesting is its relation to modern small-scale plasticity. It will be demonstrated that the discrete strain magnitudes observed in the bulk materials in the works of [4, 21] result in absolute elongations that are comparable to those associated with the large strains seen in the discrete plasticity of micron-sized samples using nano-indentation based flat-punch methods [38, 39]. This latter work has given strong evidence of critical scale-free plastic intermittency, the theories of which are generally concerned with pre-cursor plasticity prior to the flow regime. Thus, in addition to comparable plastic activity between bulk micro-plasticity and small-scale plasticity, there exists a strong theoretical connection between the two. This suggests micro-plasticity should give insight into the nature of scale-free behavior in bulk systems, and whether or not such a phenomenon is a result of the initial structure and already present in the first discrete plastic event, or that it is a result of the driven structure due to the external loading. Placing micro-plasticity in this context naturally raises the question of the origin of the small-is-stronger size effect [40, 41], and whether or not it is a manifestation of a change in statistics of bulk micro-plasticity, or a change in mechanism, due to the reduction in sample volume. The resolution of this question would then place clear limits of our proposed link between micro-plasticity and small-scale plasticity.

## 2 Early observations of bulk micro-plastic flow

*2.1. Uniaxial deformation experiments and some microstructural observations*

Micro-plastic flow and pre-yield deformation experiments were founded on the opinion that, strictly speaking, there is no elastic limit [30, 31]. This was motivated by the observation that any small increment of stress would also result in an irreversible plastic strain event [7, 8, 42]. Similar conclusions were made by various subsequent studies, some of which will be highlighted in more detail below [20, 21, 33, 43-45].

Based on the question if the yield stress is a value of fundamental significance or only a value that depends on the sensitivity of the method used to detect permanent deformation [3], Chalmers conducted a remarkable set of uniaxial deformation experiments on β-Sn-crystals with a displacement resolution of $3\times10^{-7}$ cm in 1936 [4]. It is noted that this displacement resolution is about 10-50 times less than the majority of micro-yield experiments introduced in Section 1.1. The observed permanent time-dependent plastic strain at a stress as low as 0.5 MPa was termed micro-plasticity, with evidence that the creep rate becomes zero only at zero stress or infinitely long times – observations also found for low-carbon steels [46, 47]. It was shown that a clear separation exists between the plastic strain rate (creep rate) in the micro-plastic regime and beyond the macroscopic yield stress commencing at 1.1-1.7 MPa. This finding is reproduced in Figure 2 for three different grades of β-Sn, where the applied stress is shown as a function of creep rate. Chalmers indicated the approximate transition from the micro-plastic regime (α-regime) and the macroscopic flow regime (β-regime) with the point P that marks yielding.

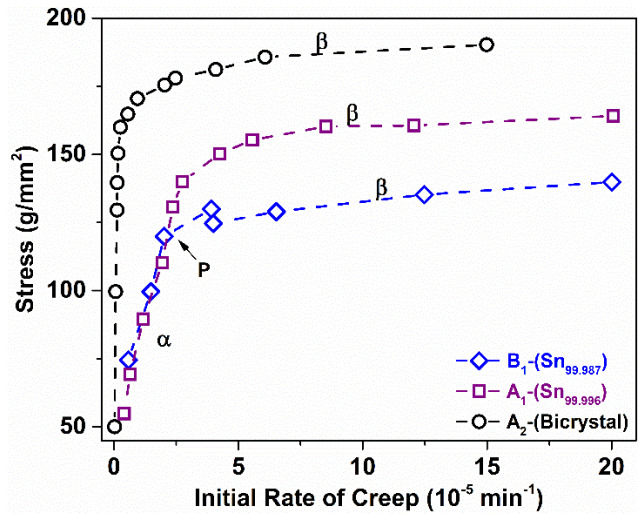

Figure 2: Stress as a function of initial creep rate for two different grades of Sn single crystals, and a Sn bi-crystal. The macroscopic yield stress in indicated with a P, separating micro-plasticity (α) and macroscopic flow (β). Reproduced from Ref. [4] with the permission from the Royal Society.

With regard to the hysteretic micro-strain loops introduced in Section 1.1, it is worth noting that such reversible non-linear strain recovery was not seen by Chalmers. He nevertheless discussed this aspect because closed-loop micro-strain hysteresis had been reported earlier by Hanson [5] for Zn loaded under torsion. The interpretation of the α-regime was based on possible discrete motion of dislocations due to thermal agitation; that is, thermally-activated collective-intermittent motion in modern terms. Since the rate of creep decreases with time for a constant applied stress in the micro-plastic regime, it was concluded that the crystal is exhausted of mobile dislocations exiting the surface. Plastic pre-yield creep in low-carbon steel concluded that the decay in micro-strain rate is rather due to an increasing back stress from dislocations generated at a source, until an equilibrium between the internal and the applied stress is reached [46]. Interestingly, Figure 2 indicates very small or virtually no change in creep rate in the micro-plastic regime, suggesting an insensitivity of the underlying dislocation motion to the applied stress level – a finding we shall relate back to in Section 4.1 when we discuss modern small-scale deformation experiments. Figure 2 also suggests that, in this case, crystal purity does not have a strong influence on the micro-plastic creep rate and that a bi-crystal geometry results in a negligible micro-plastic creep rate ($A_2$).

The numerous micro-strain experiments that mainly focus on closed hysteresis loops ((iii) in Figure 1), investigated the effects of microstructure and deformation history in great detail. It was noted by Meakin [19] that the true elastic limit, $\tau_E$, which is directly linked to the frictional stress via $\tau_E = 2\tau_F$, should reflect elastic moduli values that are measured with an independent technique and that they should also be insensitive to the initial defect structure. That was indeed the case for uniaxial quasi-static deformation of Mo, W, Fe, and Be [19, 23, 26]. $\tau_E$ was reported to be mildly dependent on temperature for iron in the range of 50-298 K [26]. Micro-strain experiments on Zn single crystals that were deformed past $\tau_{\mu p}$ (open loop behavior) revealed time-dependent recovery of plastic strain upon unloading, where the magnitude of the reverse strain increased with the total strain of prior loading [22]. This was observed at nearly zero stress (ca. 0.014 MPa, where $\tau_Y \approx 0.25$ MPa), highlighting a large micro-plastic regime. In contrast to $\tau_E$, the transition to the micro-plastic regime at $\tau_{\mu p}$ was observed to sensitively depend on the deformation history and temperature [19, 23, 26, 47] – loading once to the first detection of $\tau_{\mu p}$, and subsequently reloading to the same stress will in the first cycle show some permanent deformation, but in the second identical loading cycle loop-closure is recovered. Another series of two identical stress cycles would result in the same observation at higher stresses, which therefore is compatible with the AE-measurements mentioned earlier. Such measurement thus demonstrate the complicated interpretation of $\tau_{\mu p}$, and show that a specific deformation history may reveal that $\tau_{\mu p} = \tau_Y$. Strictly speaking, this means that micro-plasticity, as defined in Figure 1 by the range $\tau_{\mu p} \rightarrow \tau_Y$ can be beyond experimental detectability if the permanent plastic pre-strain is high enough [23]. Likewise, $\tau_{\mu p}$ may be observed close to $\tau_E$ for well annealed crystals, where closed looping does not occur [29]. In addition to $\tau_{\mu p}$ being sensitive to the deformation history of the material, there is a marked temperature effect, as was shown for Fe – as a function of decreasing temperature, $\tau_{\mu p}$ increases increasingly fast, whereas different pre-strains show a rather monotonic increase in $\tau_{\mu p}$ [26].

Most micro-strain studies did not focus explicitly on the stress regime between $\tau_{\mu p}$ and $\tau_Y$, but some studies were dedicated to the transition from micro- to macro-plasticity and the associated mechanistic changes. Relating the change in rate to the flow stress change via $\partial ln\dot{\gamma}/\partial \tau = v^*/kT$, where $\dot{\gamma}$ is the applied strain rate, $\tau$ the applied shear stress, $kT$ the thermal energy, and $v^*$ the activation volume, evidenced a distinct change in $v^*$ as a function of plastic strain for Mo single crystals when transitioning from the micro-plastic to the macro-plastic regime (Figure 3) [19]. In the micro-strain regime $v^* \approx 2000$ b$^3$, whereas 20 b$^3$ was found in the macro-plastic regime. The latter value is well in agreement with other studies on bulk bcc metals [48]. To rationalize these large changes in activation volume using the traditional bulk picture (where the activation volume is proportional to the area swept by the dislocation) becomes problematic due to the small distances associated with activated dislocation motion in the microplastic regime. Rather, one must resort to a microscopic picture in which local dislocation (atomic structural) evolution has a distinct internal stress signature that will either aid or abet the applied external load, thereby giving the associated barrier energy a stress dependence. Such a local picture of activation volume has been fully characterized in the study of the stress dependence of localized structural excitations within model glassy systems - see Section 3.3 and Ref. [49].

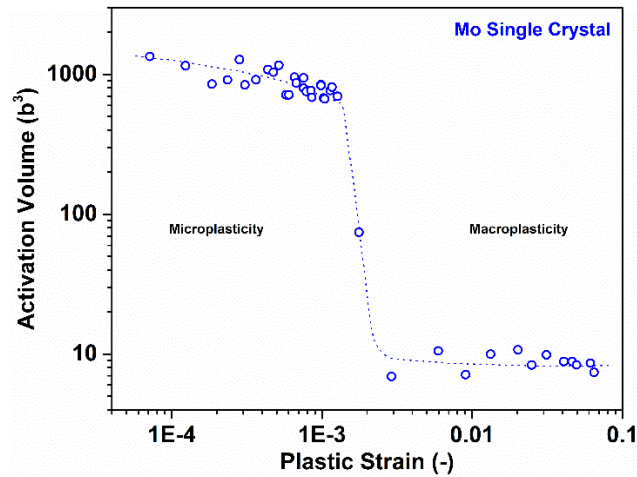

Figure 3: Activation volume as a function of plastic strain for Mo single crystals at room temperature [19]. © 2008 Canadian Science Publishing or its licensors. Reproduced with permission.

It is noted that the data of $v^*$ prior to the drop in Figure 3 is (at least partly) in the micro-plastic regime. The micro mechanistic interpretation of this behavior, a two orders-of-magnitude drop in the activation volume, and the pre-yield region in general, has for bcc lattices been ascribed to a transition from a kink-motion controlled to a kink-nucleation controlled regime [19, 26]. In other words, within the regime between $\tau_\text{E}$ and $\tau_{\mu\text{p}}$ initial dislocation motion in bcc lattices proceeds by motion of pre-existing edge kink components oriented along $\langle 111 \rangle$-directions where the Peierls barrier is the smallest. Now reversible bowing occurs in-between pinning points, consistent with the strong dependence on crystal purity and pre-strain. Above $\tau_{\mu p}$, the micro-plastic regime involves passing of pinning points by these edge components, and thus permanent plastic strain. Reaching $\tau_Y$, kink nucleation of the thus far, immobile screw components will begin to dominate, becoming the rate-limiting process for macroscopic plastic flow. This scenario does not explain the rapid hardening seen between $\tau_{\mu p}$ and $\tau_Y$, which still falls into the regime of an approximately constant $v^*$. Proposed explanations were dislocation-dislocation interactions, exhaustion of kink nucleation, and a mismatch between the plastic strain rate (and thus dislocation velocity) and the speed of the testing machine [27]. In view of the approximately constant $v^*$ in Figure 3, only the latter two proposition seem to be compatible options. This is in agreement with the conclusion that anelastic edge dislocation motion dominates the stress-strain response up to $\tau_{\mu p}$ in ultra-pure iron, after which screw dislocations begin to move [25]. In the same work, it was also deduced that the increasingly large stress span between $\tau_{\mu p}$ and $\tau_Y$, with decreasing temperature, is governed by a reduction of the cross slip rate of screw dislocations leading to significant work hardening and an increase in $\tau_Y$ with decreasing temperature in iron. In analogy to fcc metals, this strong work hardening, with slopes that are indistinguishable from the elastic behavior in a macroscopic test, is comparable to stage II hardening. Since work hardening due to screw dislocation interaction is a process occurring in the macroscopic flow regime of bcc metals, it must be concluded that the strain at which the abrupt change in activation volume occurs in Figure 3, has to be temperature dependent and may occur prior to $\tau_Y$, if the temperature is high enough and a high cross slip mobility of screw dislocations exists. Irrespective of the exact mechanistic origin of the transition in activation volume between micro-plastic and macro-plastic flow, such a result conveys that different unit mechanisms are at play for the two regimes of plasticity.

On the basis of total micro-strain magnitudes (the sum of elastic (up to $\tau_E$), anelastic ($\tau_E \rightarrow \tau_{\mu p}$), and micro-plastic ($\tau_{\mu p} \rightarrow \tau_Y$) deformation) it was shown that a power-law scaling with stress exists universally for different grain sizes in bcc, fcc, and hcp polycrystals [50]: $\gamma = C/\tau_0 \times D^3(\tau - \tau_0)^2$, where $C$ is a constant, $D$ the grain size, and $\tau_0$ the stress to activate the first source. The summarized findings are displayed in Figure 4. This data shows that all lines for different grain sizes tend towards the same value on the abscissa, meaning that independent of grain size there exists a stress below which no strain is mediated. Similar to the temperature dependent micro-plasticity measurements [25], these results also show that the magnitude of the micro-plastic region can be different for the same material with different grain sizes. Furthermore, a grain-size dependent power-law relation of type $\sigma = C + B\varepsilon_{\mu p}^n$ was reported for the stress-strain relationship in the micro-plastic regime [51, 52]. Expressed as a plastic strain rate as a function of stress $\sigma$ and dislocation density $\rho$, $\dot{\varepsilon}_\mathrm{p} = \rho b B \sigma^m$, where $b$ is the magnitude of the Burgers vector, $B$ the dislocation velocity at unit effective stress, and $m$ is the temperature dependent dislocation velocity stress exponent, a direct link between the Orowan equation and an internal length scale (defined by $\rho$) was made [45]. The scaling between $\log(\dot{\varepsilon}_\mathrm{p})$ and $\log(\sigma)$ was found to be linear, directly giving $m$. This result supports the notion that $\rho$, and thus any relevant length scale, remains approximately constant throughout the micro-plastic regime. This also illustrates that the macroscopic elastic limit should be given for a specific plastic strain rate when thermal activation is at play.

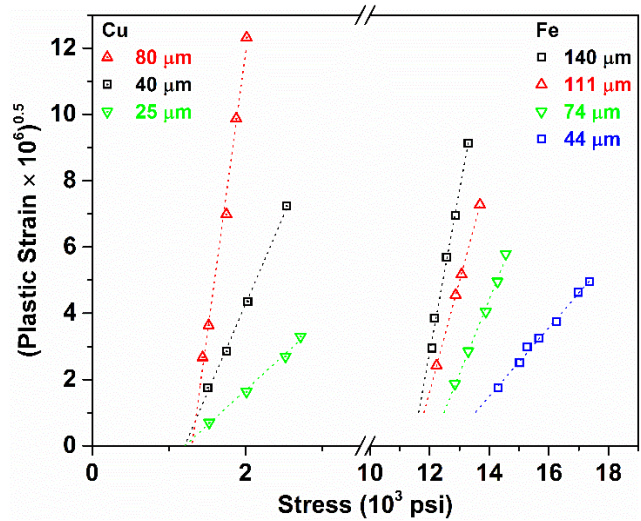

Figure 4: Square root of plastic strain as a function of stress for Cu and Fe with different grain sizes. Reprinted/reproduced from Refs. [50, 53], with permission from Elsevier.

The large body of work focusing on the stress-strain response in the micro-plastic regime has however given little direct microstructural insight. This is primarily due to the very small fraction of grains that display plastic deformation prior to yield. Probably the most straight forward experimental route to do this is to conduct etch-pit investigations, extensively used to trace dislocation movements as a result of stress pulses [54, 55]. Using this method, it was concluded that for Cu crystals with a typical dislocation density of $\sim 10^6$ cm$^{-2}$ the stress required for dislocation multiplication is higher than the stress to overcome impurity barriers, which means that no increase in dislocation density can be expected prior to macroscopic yield [56]. Silicon-iron has also been studied via etch pitting to elucidate the onset of slip-band formation and micro-plasticity [57, 58]. Investigating two different grain-sizes (20 μm and 170 μm) reveals that only 0.9% of the 20-μm grains in the gauge length deform plastically up to $0.91 \times \tau_\mathrm{Y}$ [57]. First evidence of grain yielding was detected at ca. $0.58 \times \tau_Y$ and the average number of slip lines per grain remained as low as 1-4 until the onset of Lüders' band formation. At the larger grain size of 170 μm a more gradual increase in the fraction of deforming grains was seen, amounting to 7% at ca. $0.94 \times \tau_Y$, where deforming clusters emerge. The average slip-band spacing in the micro-plastic regime varied between 500 nm and 2 μm, irrespective of grain size. Whilst the variation of lower yield stresses was very compatible with the Hall-Petch trend, it is noteworthy that individual grain yielding was insensitive to the grain size and was observed at the same stress as for a single crystal, corresponding respectively to $0.75 \times \tau_Y$ (170 μm) and $0.58 \times \tau_Y$ (20 μm). Quantitative transmission electron microscopy (TEM) in the micro-plastic regime of high-purity polycrystalline copper indicated a two stage pre-yielding, in which first dislocations were nucleated at grain boundaries and subsequently emitted into the grain interior [59]. TEM in conjunction with micro-plasticity experiments of differently deformed Ag microstructures (50 μm grain size) revealed a strong dependence of the microscopic yield stress on the degree of annealing, but the subsequent steep strain hardening response prior to macroscopic yielding remained the same [60]. This has the implication that the dislocation movement in the micro-plastic regime is controlled by a much finer scale than given by any grain size or length scale characterizing the dislocation structure. Extensive annealing practically eliminated all micro-plasticity, increasing the yield strength to even stresses above values of the cold rolled material, which was interpreted as the emergence of a nucleation dominated regime.

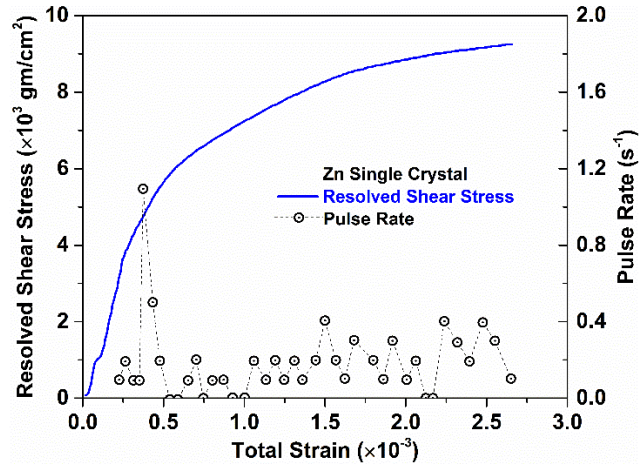

Figure 5: Resolved shear stress and AE-pulse rate as a function of total strain for Zn single crystals. Reprinted/reproduced from [9], with the permission of AIP Publishing.

AE sensing can reveal the plastic activity prior to $\sigma_Y$. The AE signal could be decomposed into a continuous component and a burst component [11, 61, 62]. Historically, the continuous signal was associated with elastic wave energies originating from dislocation motion, whereas the burst component was generally associated with more abrupt structural changes such as internal crack nucleation and propagation, or violent collective dislocation rearrangements now referred to as avalanches. Whilst such experiments were initially rare, some early studies clearly showed that a significant number of pulses may occur in the micro-plastic regime or at strains that nominally are well below the offset strain for macroscopic yield. Both irradiated LiF and Zn single crystals exhibited a peak in pulse rate prior to $\tau_Y$ [9]. AE-pulses were recorded upon the application of a load, reached a maximum pulse rate in the micro-plastic regime, and decayed quickly at the transition to macroscopic flow. Figure 5 displays this for Zn single crystals. Similar observations where made for Cu single crystals and Al polycrystals [11], with the additional insight that AE-pulses where absent after unloading and subsequent reloading until a stress larger than the previous peak stress was reached. This feature is generally known as the Kaiser effect [12, 15], and interpreted on the basis of mobile dislocation exhaustion in crystals. At the transition to macroscopic flow, a continuous AE-signal, with a rate-dependent pulse amplitude, is initiated. The continuous part of the AE-signal was for most fcc metals the dominant contribution, which Rouby and co-workers [63, 64] describe as the uncorrelated superposition of events coming from different sources. Such a picture is very much compatible with early AE experiments on metallic single and poly-crystals in the flow region [65]. This work revealed plastic flow also occurs discontinuously at a frequency (AE pulse signal frequency) which is proportional to strain rate, with corresponding plastic strain magnitudes estimated to be $\sim 10^{-7}$. Such small strain increments suggest that slip-bands originate from a large number of such events. This work is notable by its usage of the micro-plastic terminology for the flow region, suggesting that when discrete plasticity can be detected – *where ever* it might occur in the stress-strain curve – the concept of micro-plasticity can be embraced.

*2.2 Torsion – intermittency during micro-plastic flow*

Higher strain resolution than that achieved in the above mentioned works addressing the micro-strain phenomenon could be achieved in torsion experiments of both polycrystalline Cu and single crystalline Zn [20, 21, 33]. With a strain resolution of $10^{-9}$, mm-sized Cu samples yielded at stresses of 0.02 MPa [20] and Zn at 0.01 MPa [21, 33]. Figure 6 displays the very early onset of loading ($\gamma_p < 4 \times 10^{-6}$) for a Zn-crystal [21, 33], where the inset highlights the intermittent plastic response in the micro-plastic regime. Discrete strain events with magnitudes of the order of $10^{-9}$ - $10^{-7}$ were found to be fully plastic [20]. It is noted that the smaller strains are equivalent to a sub-nm displacement resolution, and thus comparable to today's nano-indentation systems, meaning that these torsion experiments resolved the displacement activity equivalent to individual Burgers vectors – for bulk systems. Furthermore, the burst size varies across two orders of magnitude, a fact that is not necessarily compatible with a unit mechanism and an aspect which will be discussed in Section 4.2 (Scale-free slip behavior). In the work of Tinder and co-workers, the resolved small plastic segments were separated by regions identified as elastic deformation. Indeed, the derived elastic modulus was comparable to that obtained from ultrasonic measurements indicating the high-quality of the deformation experiment.

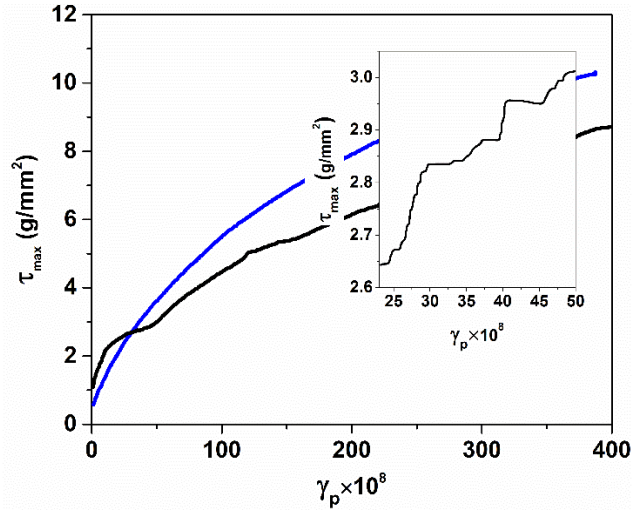

Figure 6: Stress as a function of plastic strain for Zn single crystals deformed via torsion. The inset shows details of the intermittent micro-plastic activity found at all plastic strains. Reprinted/reproduced from Ref. [21] with permission from Elsevier.

Estimating the length of a possible Frank-Read source with $L = 2\,Gb/\tau \approx 3$ mm, it was concluded that the observed strain bursts are not due to such a source mechanism, because the obtained length was both larger than the specimen and the subgrain size of the material. Furthermore, dislocation breakaway from impurity clouds was excluded, leading to the interpretation that it is the small mobile fraction of a grown-in dislocation network, which can trigger sudden progressive rearrangements with little change in the total dislocation line energy. Consequently, the rich initial intermittent plasticity in the micro-plastic regime can only be due to the mobile dislocation network. Repetitive loading and work hardening of the crystals ultimately led to the disappearance of the burst activity. This is in agreement with the earlier mentioned peaking of pulse-rate prior to macro-yield during AE-sensing [9], and eventual transition to a continuous component upon reaching stage I macro-plastic flow. Very little was, and still is, known about what fraction of an initial network is contributing to experimental pre-yield plasticity, but the fact that only a small fraction of the dislocation network is mobile was also deduced from earlier AE-measurements [11].

*2.3 Micro-plasticity in cyclic loading, internal friction and mechanical spectroscopy*

Micro-plasticity is a phenomenon that overlaps with the domain of internal friction, which can be easily understood when considering that the closed-loop micro-strain experiments addressed in Section 1.1 are essentially an internal friction experiment conducted at a very low frequency. The internal microstructural damping, $Q^{-1}$, is directly proportional to the energy loss of one cycle, $W$, relative to the global energy, $W_G$, applied to the system: $Q^{-1} \propto W/W_G$. A topic relevant to micro-plasticity is how $Q^{-1}$ might change in the strain regime below $\tau_Y$, a schematic trend of which is depicted as an inset in Figure 7.

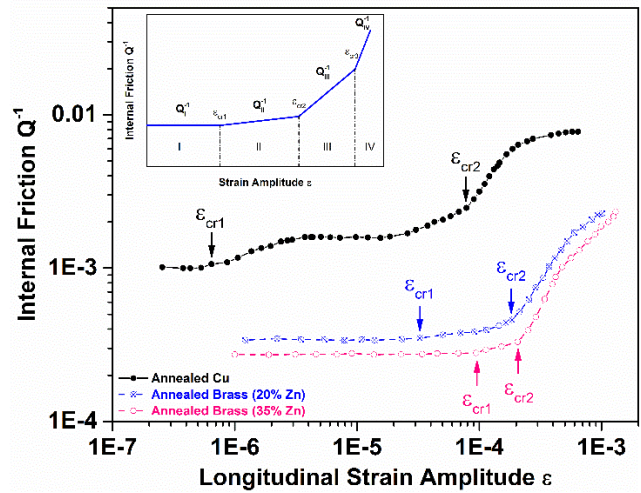

Figure 7: Internal friction versus longitudinal strain for annealed oxygen free Cu, and two different grades of annealed brass. Reprinted/reproduced from Ref. [66], with the permission of AIP Publishing. The inset displays schematically different regimes of internal friction, as proposed by Puškár [28].

In regime I of the inset in Figure 7, the strain amplitude is too small to induce dislocation motion, and it is only when the strain amplitude exceeds $\varepsilon_{cr1}$, inelastic interaction of dislocation segments with obstacles becomes detectable. This is the strain amplitude regime in which, at a given temperature and constant frequency, the Bordoni [67] or Niblett-Wilks [68] maxima in $Q^{-1}$ can be found. These maxima indicate dislocation relaxation, allowing to assess thermally activated dislocation motion over intrinsic potential barriers, and have been used to study fcc, bcc, and hcp lattices as a function of alloying concentration, impurity content, pre-straining etc [69-72]. Essentially, the size and position of the internal friction peaks depend primarily on the amount of pre-deformation of the sample [73, 74]. Dislocation relaxation spectra therefore focus on structural constancy (no plastic deformation) during the measurements and can be placed in a strain amplitude regime outside that of micro-plasticity. The mild strain dependence of $Q^{-1}(\varepsilon_{cr1} < \varepsilon < \varepsilon_{cr2})$ was successfully described by Granato and Lücke [75, 76] via an internal damping $Q_m^{-1} = f(L_p, L_n)e^{f(L_p)/\varepsilon}$, where both $L_p$ and $L_n = n \times L_p$ are internal length scales signifying the dislocation segment length between pinning points and the combined length of several such segments lengths, respectively. In this picture, the dislocation line may generate a loop at the critical strain $\varepsilon_{cr2}$ and therefore act as a source if $L_n$ is sufficiently long and the external strain amplitude large enough. Passing $\varepsilon_{cr2}$, micro-plastic deformation takes place that has a saturating character, since the plastic damping $Q_p^{-1}$ attains a stabilized value. Here it is assumed that both the density and distribution of dislocations no longer change significantly before reaching the transition to macroscopic yield [77]. In fact, it was found that only the density of mobile dislocations increases in the micro-plastic regime. At a strain amplitude of 0.2%, the relative growth of the total dislocation density in a CuAl-alloy was found to amount to ca. 10% [28]. Subsequently, when strain amplitudes are larger than $\varepsilon_{cr3}$, damage accumulation as in conventional fatigue sets in, and the damping becomes a function of the loading cycle. For iron of different grain sizes, it was shown that the transition from regime II to III increases with decreasing grain size [78]. The same body of work also reports the activation volumes for both fcc and bcc crystals. Those for bcc are found to be very much in agreement with Figure 3 and also a dramatic drop in $v^*$ was observed at a given plastic strain [78, 79]. However, the values of $v^*$ in regime III are still high when compared to macroscopic plastic flow, as is exemplified with polycrystalline Fe where in regime III a grain-size independent activation volume of some hundred $b^3$ is observed.

The aforementioned strain amplitude regimes are well exemplified with the data on high purity polycrystalline Cu and brass from Ref. [66] shown in Figure 7. After reaching a strain amplitude of $\varepsilon_{cr1}$ the internal friction increases, but clearly $\varepsilon_{cr1}$ and the extent of regime II is strongly dependent on the microstructure. It is seen for the case of Cu that a change in internal friction occurs two times prior to reaching the onset of micro-plastic deformation at $\varepsilon_{cr2}$. At this point, it was postulated that Frank-Read sources begin to operate and that long range motion of mobile dislocation segments begins [66]. Clearly seen in the case of Cu, the associated damping again levels off towards a saturation value of $Q^{-1}$, whilst still in a strain amplitude regime below $10^{-3}$. A similar plateau may be reached for the two types of brass, but is not captured in Figure 7 or the original data. Also not evident in Figure 7 is the strong but non-trivial dependence of $Q^{-1}(\varepsilon)$ on the amount of macroscopic pre-straining, of which the detailed mechanistic origins remain an unsolved discussion around specific dislocation mechanisms [80, 81]. Even though many of these mechanistic questions are to date still unresolved, the body of data supports the view of small microstructural changes in the pre-yield regime. Clearly, the foregoing sentence needs clarifying work, where the focus should be to elucidate if the early microstructural development is statistically significant nor not? That is, do the macroscopic parameters that define the initial structure change non-negligibly?

*2.4 Some comments on classical observations of micro-plasticity in crystals*

The few selected classical works on micro-plasticity and pre-yield observations barely make justice to the large amount of available literature. The purpose of highlighting a few main themes is to set the stage for the developments and discussion following in Section 4. In particular, the classical works clearly focused on particular mechanisms that dominate in particular stress or deformation regimes, as exemplified with the data presented in Figure 3, and the discussion around Figure 7. The same applies to the power-law description of the pre-yield stress-strain behavior that is implicitly based on well-defined mean quantities. The continuous AE-signal superpositioned with bursts also justifies some governing process centered around a defined mean. These approaches have been very successful in describing flow by homogenizing the underlying detailed dislocation processes and to treat plasticity as smooth (laminar) flow.

Early deviations from this paradigm are the few observation of pre-yield intermittent deformation. In particular, the discussed work in Section 2.2 contains, as will become clear in Section 4, remarkable indications for what today is contained in the view that plasticity is a process that exhibits some scale-free properties within certain bounds. Indeed, Tinder's work [21] reveals strain-burst magnitudes covering two orders of magnitudes, with increasing values upon approaching macroscopic yield (criticality? - see Section 4.2). Thus, there seems to be no well-defined length scale associated with the observed strain bursts. A lack of such a length scale also suggests a less local mechanism and diverging correlation lengths, both of which are at odds with the description of smooth plastic flow based on unit mechanisms [82]. Furthermore, the fraction of dislocations taking part in micro-plasticity was, if investigated, always found to be small, and no evidence of significant changes to the pre-existing network has been reported during micro-plastic flow.

**3 Micro-plasticity in metallic glasses**

*3.1 What is micro-plasticity in metallic glasses?*

Defining micro-plasticity for metallic glasses is less straight forward than for crystals. This is because the atomically disordered structure of a glass allows for local plastic transitions at any stress. The reasoning for this can be found in the expected broad distributions of barrier energies and attempt rates, if one considers thermally activated plastic events. Therefore a true $\tau_E$, as defined for crystals should not exist, even though the assessment of deviations away from a linear elastic response at low stresses is experimentally very challenging. In fact, macroscopically, metallic glasses exhibit a very robust elastic response and reproducible yield stress, with Weibull moduli that can be as high as for steels under uni-axial compression [69]. This does not necessarily imply a lack of structural activity prior to yield, but rather suggests a lack of strong influence of stress in breaking the symmetry of the system [49]. Indeed, there is now growing evidence of irreversible atomic rearrangements even well below the macroscopic yield stress, clearly revealing rich micro-plastic activity prior to the formation of shear bands. It is noted that some reports on in-situ AE-measurements show distinct pulses prior to $\tau_Y$ [13, 14]. These should be due to (non-system spanning) shear-band formation as a result of a more complex (not uniaxial) loading geometry and are not included in our consideration of micro-plasticity.

Micro-plasticity, defined as any plastic activity in the inhomogeneous flow regime that does not involve shear banding may be appealing, but ignores the fact that shear bands can display creep, especially when the shear-band relaxation dynamics is quenched at lower temperatures [83, 84]. We will nevertheless take this view. The precise atomistic mechanisms that describe local plastic activity according to this view are best captured by the general terminology of dissipative rearrangements, which are either purely mechanically driven (athermal) or thermally activated instabilities. At the atomic scale such processes have been described by a set of different concepts, including shear transformations [85, 86], flow defects [87], or via a hierarchy of smaller relaxation processes (β-relaxations) mediating larger events (α-processes) conceptualized in a potential energy landscape [88, 89]. For the details of these, we refer to the elegant reviews recently devoted to the mechanical behavior of metallic glasses [90, 91]. In the following, we will focus on experimental indications of micro-plasticity in the nominally elastic and inhomogeneous deformation regime, and the valuable insights gained from modern atomistic computer simulations.

*3.2 Micro-plastic behavior of metallic glasses in experiments*

Resolving small irreversible processes of plastic dissipation at small stresses or in-between shear-band activity is experimentally very difficult, and can only be observed with strain resolutions typically available in small-scale experiments or with methods that directly probe atomic structure. Noticeable plastic deviations away from linear elastic tensile loading was, for example, observed for Co-based metallic glass micro-wires [92, 93]. Passing a specific stress level, the wires exhibited marked irreversible plastic deformation of the order of 0.6%, with repeated load-unload cycles giving open loop behavior, similar to case (v) in Figure 1. Subsequent positron annihilation lifetime spectroscopy indicated an increased signal for interstitial holes and sub-nanometer voids as a function of loading. Additional estimates based on average shear-transformation-zone (STZ) sizes or barrier energies indicate a strong reduction in both quantities with decreasing wire diameter, and in comparison to bulk samples [93]. This could furthermore be supported by investigating the barrier energy distribution of the material, which was reported to shift to lower peak values and substantially narrow with sample size reduction [92].

Cyclic or quasi-static compressive loading of micron-sized metallic glasses also reveals a strongly inelastic signature. During cyclic loading in the elastic regime it was observed that the inelastic response increased with applied stress rate, but that all strain could be recovered after unloading [94]. Similar observations stem from cyclic tensile loading of macroscopic samples [95]. Whilst increased anelastic responses were measured with increasing stress-rate during cycling, quasi-static micro-compression revealed increasing irreversible plastic strain with decreasing stress rate [96]. Clearly, there are different rate-effects that seem to reveal micro-plastic activity, some of which can be understood with viscoelasticity that originates from a distribution of softer and stiffer regions in the glassy matrix [95].

Probing directly the atomic-scale elastic strains with high energy x-ray diffraction [97], it could be shown during in-situ loading past the elastic limit, that the linear elastic strain relation breaks down at about $0.77 \times \tau_Y$ [98]. The examination of changes in diffraction peak-width revealed different behavior for different metallic glasses, and it was argued that a redistribution of free volume to a more homogeneous state occurred. It is noted that such early deviations from linear stress-strain relationships have so far only been observed for specific alloys [98, 99], likely to be those that exhibit some amount of compressive ductility. This means there is either some, yet not well experimentally characterized, structural property, which mediates micro-plasticity and that also may be a prerequisite for macroscopic malleability, or it remains unresolved for other alloys systems due to a lack of measurement sensitivity.

In order to circumvent the strain resolution problem associated with quasi-static macroscopic loading, constant force measurements can be done. In such experiments, the material is loaded to a given stress for a certain time, the strain is monitored, and at a given time the sample is unloaded to zero stress. This testing protocol resolves accumulated plastic strain as a function of time, and has clearly revealed homogeneous pre-yield plasticity [100-104]. In such experiments viscoelastic deformation during loading at a given fraction of $\tau_Y$ can lead to several tenths of a percent non-recoverable strain. Figure 8 displays a typical strain-time trace for a $Ni_{62}Nb_{38}$-glass [103].

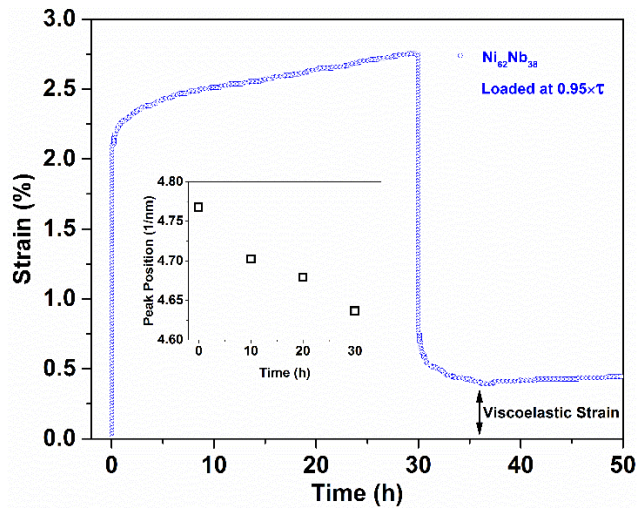

Figure 8: Strain as a function of time during constant load experiments (elasto-static) at 95% of the yield strength. Significant permanent plastic strain is seen upon unloading to zero stress after 30h. The inset displays the peak position obtained from selected area diffraction in transmission electron microscopy as a function of loading time. Reprinted/reproduced from Ref. [103] with permission from Elsevier.

Based on shear-stress induced free-volume generation, the constant strain rate has been directly related to the accumulation of excess free volume $v_f$ [105]: $\frac{dv_f}{dt} = C\tau \frac{d\gamma}{dt}$, where $C$ is a material dependent constant. Electron diffraction experiments, differential scanning colorimetry, and high strain rate molecular dynamics (MD) simulations support the view of an elasto-statically induced disordering (see inset in Figure 8, electron diffraction data), leading to a density decrease, a decrease in $\tau_Y$, and an increase in the stored heat of relaxation. Recent work focusing on the energy storage imparted via elasto-static loading also reported the same effect in tension at a remarkably low load of 1% of the yield strength [106]. Interestingly, the elasto-static testing protocol can also yield opposite results, termed as mechanical annealing in reference to the analogous temperature driven relaxation [107]. This, less frequently reported, response is in terms of a strengthening effect somewhat similar to what is also known as stress-induced hardening, where a given elastic pre-load or particular multi-axial loading state can induce relaxation of the glassy structure [108]. Similar conclusions have been drawn from permanent homogeneous deformation in bending experiments [109]. Not ignoring the results of Ref. [107], one may come to the conclusion that the mechanism underlying micro-plastic deformation during elastic pre-yield loading can be of two opposite types (rejuvenation vs ageing) due to the different stress states. However, the overwhelming evidence speaks for micro-plastic shear-driven rejuvenation. One interesting aspect of these findings is that stress and temperature protocols are expected to bias the underlying atomistic barrier energy landscape differently, as it is known that a load leads to a broadening of the initial barrier-energy distribution [49, 110], but the application of a thermal protocol does not. Therefore, it remains a topic of future efforts to uncover the origins of such opposing results obtained with elasto-static loading.

Due to the thermally activated nature of plastic flow of metallic glasses at all temperatures, increasing the temperature above room temperature (or more precisely to a higher fraction of the glass transition temperature) naturally amplifies permanent plastic pre-yield strains. Under such conditions, stress-relaxation has been shown to result in increasing amounts of micro-plastic strain when approaching stress levels comparable to the macroscopic yield [111]. This can be rationalized by a gradual transition from sampling primarily β-relaxation modes until the increasing amount of local structural excitations merge to mediate α-transitions and therefore macroscopic flow. Both α- and β-modes are also assessed during dynamic mechanical analysis (DMA), which has been used intensely to investigate structural relaxation at a given applied alternating load and frequency. During cycling, the temperature is increased, leaving a clear signature of pronounced relaxation dynamics at given temperatures for some alloys systems. In particular, La-based and Pd-based glasses exhibit a broad low temperature peak, typically associated with β-relaxations [112]. The strong structural activity has in some cases also been linked directly to the emergence of tensile ductility [113]. From the perspective of micro-plasticity, such measurements are relevant as they probe structural modifications that are not only temperature-induced relaxations, but also stress or strain driven. In fact, with the increasing body of evidence for vast plastic activity at low stresses in the pre-yield regime, it remains unclear to what extent such DMA experiments probe the as-cast reference state of the material, or if a continuously deforming out-of-equilibrium structure (change of reference state) is probed while ramping the temperature.

Using DMA to conduct a constant force experiment (thus the same as an elasto-static test), and by analysing the load- and temperature-dependent time the material needs to undergo a given increment in length, a change in structural plastic dynamics has been evidenced [114]. Figure 9 reproduces selected results of this waiting time analysis taken from Ref. [114]. The data exhibits a two-regime power-law scaling in the statistics of waiting times depending on which part (creep times $t < 100$ min, or $t > 100$ min) of the creep curves is considered. This indicates that both above and below the cross-over no specific mean waiting time defines the underlying process. The trend in Figure 9 was interpreted as a cross-over from random three-dimensional plastic activity to cooperative two-dimensional activity and therefore marks the transition from micro-plasticy to shear-banding and thus inhomogeneous macroscopic plasticity.

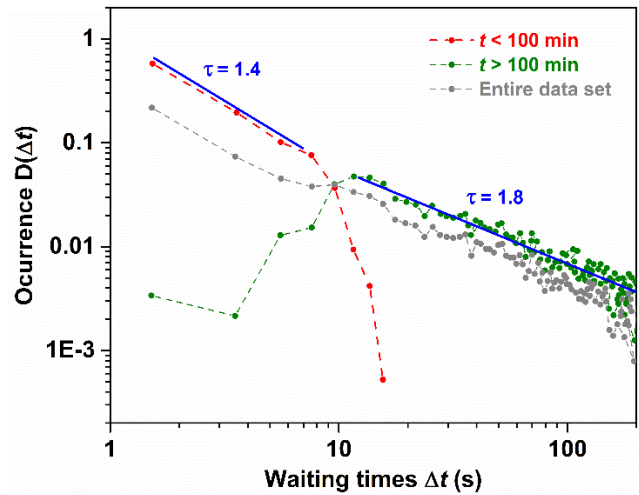

Figure 9: Two-regime waiting time ($\Delta t$) distribution obtained during creep measurements at 320°C of a Pd-based metallic glass ribbon. The data with $\tau = 1.4$ represents the first 100 min of the creep curve, and the data with $\tau = 1.8$ creep-curve times larger than 100 min. Data reproduced from Ref. [115].

*3.3 Micro-plastic behavior of metallic glasses in simulations*

Atomistic simulation allows the possibility of studying the mechanisms of local plasticity with microscopic resolution. However, because of the limited physical time simulated, strain rates many orders of magnitude larger than in experiment are used to achieve non-negligible plastic strain magnitudes. Thus the type of plasticity studied by atomistic simulation is mainly athermal, manifesting itself as mechanically driven local instabilities of the atomic structure. Despite this caveat, such simulations have provided extremely valuable insights into what can happen at the atomic scale. A detailed overview is found in Refs. [90, 116]. The main finding of atomistic simulations is that athermal plasticity is mediated by spatially local plastic transitions, which collectively result in the macroscopic plasticity of (athermal) shear banding [117-125] – a macroscopic flow mechanism that experimentally is known to be thermally activated [126]. A common question addressed in such simulation work is what local structural features lead to a particular region becoming structurally unstable? The search for such a correlation with local environment was motivated by the concept of liquid like regions [127], and regions within the glass structure predisposed to driven plastic activity were considered in detail by Demkowicz and Argon [124, 128, 129]. Very recent work by Falk and co-workers [130] demonstrates a broad range of environments susceptible to a driven material instability characterized by a coarse-grained local yield stress that follows a low-stress asymptotic power-law distribution, whose exponent depends on the degree of relaxation of the amorphous structure.

To study local structural excitations under zero stress or at a stress low enough not to induce structural instabilities, potential energy landscape exploration methods are used. One such method is the activation-relaxation technique (ART [131, 132]), which can find the nearby saddle point configurations separating one local minimum and another neighboring local minimum. This approach has demonstrated the corresponding structural transformations are always spatially localized involving between one and up to ten central atoms [49, 122, 123, 133, 134]. The energy difference between one local minimum configuration (the starting state) and a nearby saddle-point (the activated state) gives the barrier energy corresponding to the associated structural transformation. When many such energy barriers of a given computer generated amorphous structure are binned, a distribution is found that peaks at a particular barrier energy and appears to go to zero for small barrier energies [122, 133-135]. This latter feature mainly occurs for sufficiently relaxed computer generated glass samples.

Detailed atomic scale inspection of these local structural transformations revealed them to contain a central group of atoms which undergo maximum displacement surrounded by a field of much smaller displacements that help to accommodate the central structural change [122, 133, 134]. In Ref. [134] these central structures, referred to as local structural excitations, where characterized by a chain-like geometry in which a sequence of neighboring atoms shifted to adjacent positions (Figure 10a). In the work of Fan et al. [133], such displacement magnitudes could be correlated quadratically with the corresponding barrier energy suggesting a direct link to the Eshelby inclusion construct [136], whereas in the work of Swayamjyoti et al. [49], the barrier energy could be linearly correlated with the change in internal pressure arising in the system transiting to the saddle-point configuration. Such microscopic data could, in principle, give insight into structural excitations driven by thermal fluctuations. This aspect was studied in Refs. [123, 137] using harmonic transition state theory to obtain estimates for the transition rate associated with the system passing through each saddle-point configuration obtained from ARTn. By performing finite temperature simulations as a function of increasing temperature, the static string-like structural excitation found by ART in Refs [49, 134] (Figure 10a) were also seen in dynamic simulations, where at low temperatures ($T < 0.5T_G$) the central displacements were seen to be only a fraction of the bond length [138]. However, at temperatures approaching the glass transition regime string-like excitations that involved bond-length displacements became increasingly numerous.

How do such barrier energies respond upon the application of an external load? Ref. [49] found that the corresponding barrier energy distribution shifts to lower/higher values for a tensile/compressive load, whereas for a pure shear geometry the barrier energy distribution broadens. For the case of a pure shear, inspection of individual saddle-point configurations revealed either an increase or a decrease (hence the broadening of the overall distribution) depending on the internal shear stress signature of the saddle-point configuration prior to loading. These trends are high-lighted in Figures 10b and c, the latter of which also shows the scale of the corresponding activation volume. The consequences of such a result have been investigated for a thermal-activation glass-plasticity model that assumes a distribution of barrier energies [110, 139, 140]. In this theoretical work, the emerging effective (average) barrier energy can have a quite complex temperature and stress dependence that differs markedly in the low temperature and high temperature regimes of the amorphous solid due to a shift from low temperature extreme-value-statistics to a high temperature statistics of the most probable.

The activation volumes obtained by ARTn and displayed in Figure 10c, are significantly smaller than that found in nano-scale deformation experiments [105, 131-133]. This difference can have a number of origins. Firstly, the localized structural excitations found by ARTn are clearly microscopic in scale, whereas experiments most likely probe collective activity consisting of many such plastic events and therefore best describes a shear processes via α-transitions. If instead the distribution of activation volumes related to the onset (stress) of shear-band initiation is derived [141-143], a more compatible number of a few atoms triggering shear banding is obtained. Secondly, the structural transitions obtained by ARTn generally have barrier energies that are well above $k_B T_G$, and are thus unlikely to be activated via thermal fluctuations. Indeed, experiments will most likely probe only the low energy tail of the barrier energy distribution. Recent simulation work has found that low barrier energy ARTn obtained transitions are generally less localized for less relaxed samples [133], a result which is compatible with the dynamical simulations presented in Ref. [130].

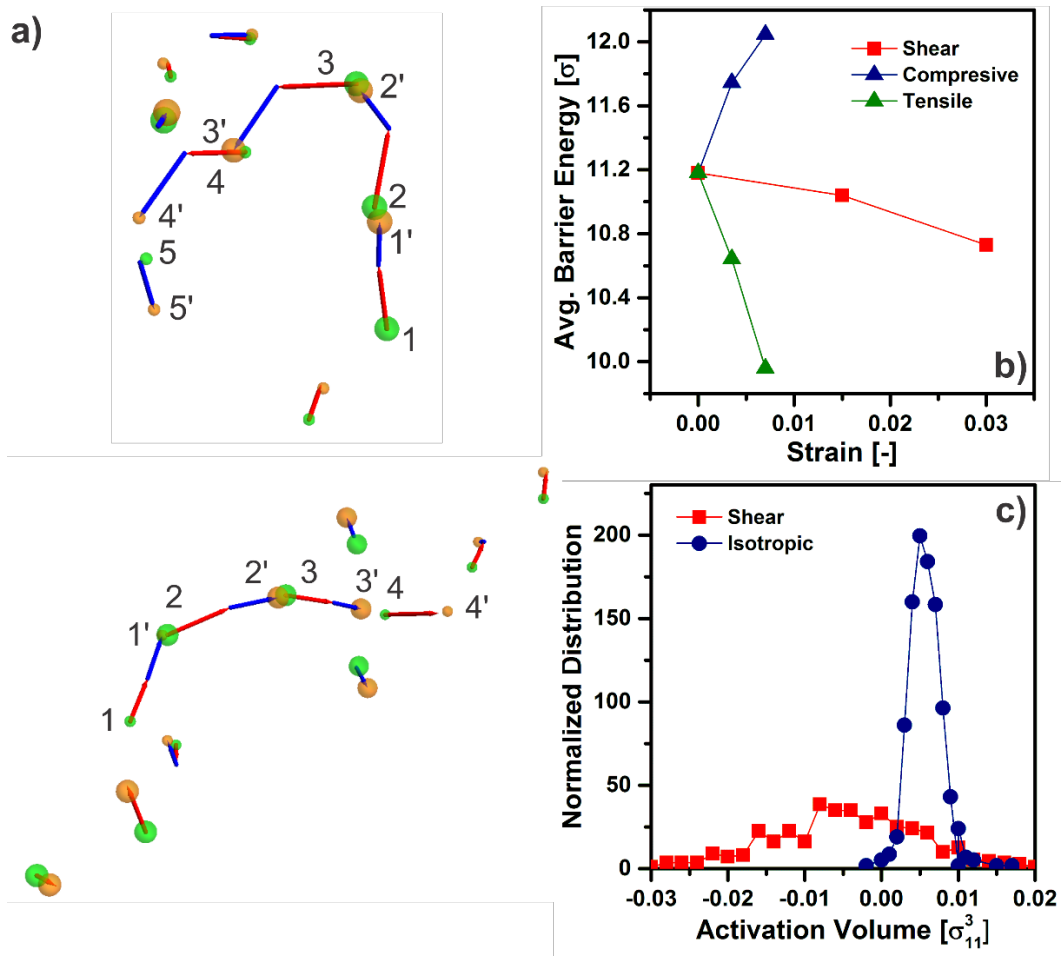

Figure 10: a) Atomic displacement maps representing two typical local structural excitation. In both cases, the initial atomic positions are visualized by green balls and the final ones by orange balls. The atomic displacements from the initial to activated and activated to final states are visualized by red and blue arrows, respectively. b) Average barrier energy for a model glass strained under different loading geometries (here $\varepsilon$ represents the Lennard-Jones energy parameter), and c) the normalized activation volume distributions due to pure shear and isotropic loading (here $\sigma_{11}$ represents the typical length scale of the nearest neighbor bond and is the Lennard-Jones length parameter). Reprinted with permission from Refs. [49, 134]. Copyright (2014, 2016) by the American Physical Society.

*3.4 Some comments on observations of micro-plasticity in metallic glasses*

Defining micro-plasticity for crystalline materials was based on the fact that plastic events only occur when $\tau > \tau_E$, irrespective if this can be resolved through macroscopic testing or not. The currently available data on metallic glasses indicates strong micro-plastic activity at virtually any stress and therefore argues for no true elastic limit. Structural changes at very low stresses were first recently recognized [106] and hint on the importance of micro-plasticity in metallic glasses as a route to modify/tune the enthalpy state. These insights lead to the exciting question how to reconcile this structural activity at any elastic strain with the robust macroscopic elastic behavior. Frequently, this very robust elastic response and the small rate-sensitivity at room temperature motivated many studies of athermal plasticity, which is in direct contradiction with time-dependent structural evolution in the micro-plastic regime. Indeed, the foregoing sub-sections clearly emphasize the thermally activated plastic behavior at any far-field stress. From a certain perspective this is unsurprising, since unlike a crystal, which has well defined defects with respect to the lattice, a structural glass admits no obvious finite lower bound in length and energy scale associated with its structural excitations.

In comparison to crystalline micro-plasticity we are only now beginning to understand how to promote and characterize local excitations in the pre-yield regime of metallic glasses [144]. How underlying length-scales may change, how the structure evolves in terms of enthalpy state under different stress states, and what mechanisms are at play, will be questions of future research. Based on recent findings evidencing the reduction of structural correlation lengths during thermal relaxation [145], one may speculate that structural disordering and therefore micro-plastic activity during elastic loading is of the opposite effect – that is, correlation length-scales would increase. Whilst structural rearrangements during micro-plasticity of crystals always involve some long-range interactions and a collective dislocation response, it remains unclear if micro-plastic structural excitations in metallic glasses have long-range interactions and potentially are correlated events. The emerging understanding of scale-free (micro)-plasticity in crystals (see next section) is timely, as it should direct us to critically investigate if such collective structural excitations in glasses also display power-law statistics known to characterize (micro)-plasticity in glasses. Finally, we note that the vast micro-plastic activity in metallic glasses has an interesting implication for the still inconclusive, but numerously reported, finding that some alloys exhibit nominally homogeneous deformation at the small scale [146-151]. Instead of seeking the origin of such size-dependent homogeneous deformation in shear-band energy criteria [146, 147] or intrinsic length-scale effects [96], it may be the density and rate of local structural transitions admitted by the different small-scale glassy structures that determines the size-effect in deformation mode.

**4 Discrete plastic flow – from well-defined mean quantities to scale-free plasticity**

Figure 1 schematically displays discrete plastic behavior, a phenomenon reported as early as 1932 by Orowan and Becker for macroscopic flow [31]. This early observation most probably played a determining role in the classic dislocation theory work done by Orowan only a few years later [34, 152, 153]. Following Orowan and Becker, numerous works have studied intermittent macroscopic deformation, and the important contribution by Tinder and co-workers [20, 21, 33] (Section 2.2) revealed that discrete plastic flow separated by elastic segments is vastly present in the pre-yield regime.

The discreteness and stochasticity of such plasticity was further substantiated by AE-experiments discussed in Section 2.1 with, as pointed out by Zaiser [154], early work focusing on the continuous AE-component. Indeed for many fcc metals, this was the dominant contribution to the AE-spectrum, starting with a maximum amplitude prior to the onset of macroscopic flow (or at yield) and decreasing as a work hardening regime of plastic flow is entered [9, 61]. These experiments revealed a very large number of energy bursts localized in time, whose statistics could be well described by a Gaussian distribution. This gives the important (and via the central limit theorem, universal) result that a well-defined energy scale associated with the AE-signal exists, supporting the idea of a statistically meaningful local average in terms of length and stress. Naturally the concept of a unit-plastic mechanism (or mechanisms) emerges, which has been the focus of many classical (micro-)plasticity treaties. This viewpoint underlies the work of Orowon, Taylor, Polanyi, Nabarro and Eshelby, and therefore a considerable part of our material science understanding of crystalline systems.

*4.1 Discrete plastic flow in nano-indentation and micro-deformation*

The continuing desire to understand bulk behavior by investigating specific defect mechanisms, has also lead to a significant decrease in length scales at which a material is plastically probed. This is well manifested in the large body of work on nano-indentation and the more recent development of testing protocols that allow deformation experiments to be performed on micro- and nano-sized specimens. Of particular relevance to this discussion is the numerously reported intermittent flow response in both nano-indentation and also small-scale testing (primarily micro-compression). In the case of nano-indentation, "pop-in" behavior has attracted a lot of attention, being for example reported for single crystals [155-159], nanocrystalline metals [160], and metallic glasses [141, 142, 157, 161, 162]. In all of these studies, the first pop-in is associated with the onset of plasticity, which may have its origin in either homogeneous dislocation nucleation, the activation of a pre-existing defect network, or (for the case of glasses) shear-band initiation. Naturally, the focus of the first pop-in has been on the critical stress at which it occurs, and also on the changes in cumulative distributions upon rate- [141, 157, 163] or temperature-dependent [164] testing. In crystals, where a pre-existing microstructure with a well-defined correlation length exists, it is the interplay between the indenter tip size and these length scales that will determine if dislocation nucleation or activation of pre-existing dislocations occur [156, 165]. Assessing for different loading rates or temperatures the statistics of the critical stress at which such plastic events occur can give mean-quantities such as activation volume or activation energy. So far, such work has mainly focused on the onset of plasticity and not on a general study of intermittent flow, which not only is captured by the first pop-in, but also by the stress-scale of subsequent instabilities at larger indentation depths.

Small-scale straining experiments, such as micro-compression, also display intermittent plastic flow. Here, stress-strain curves are obtained, which are qualitatively similar to those found in the early work by Becker, Orowan, Schmid and Tinder. Figure 11 shows the comparison between a modern micro-compression stress-strain curve (Au crystal) and a tensile stress-strain curve for a bulk Zn crystal obtained by Schmid and Valouch almost 100 years ago [30]. Even though the latter experiment has a very low stress and strain resolution, as seen by the sparse number of data points, the flow curve clearly displays strain jumps and stress drops.

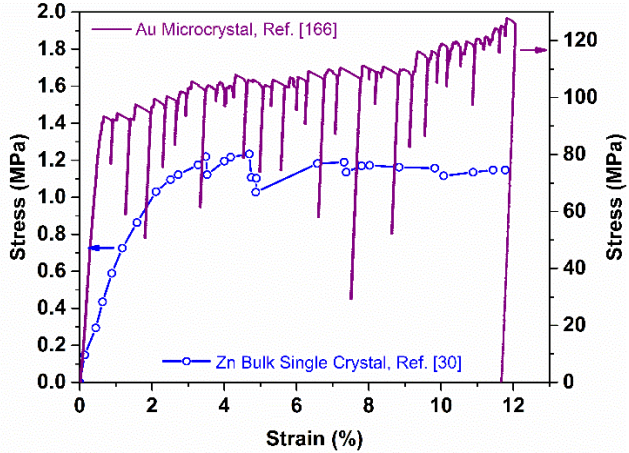

Figure 11: Stress-strain data from a Zn bulk single crystal deformed in tension in 1932 from Ref. [30] (reproduced with permission of Springer), and a micro-pillar compressive stress-strain curve from Ref. [166] (reproduced with permission from Elsevier).

The important difference in the displayed data is that the strain jumps in the macroscopic curve reflect absolute slip sizes of ca 5-10 μm, whereas the micro-crystal experiments exhibits slip sizes of a few Burgers vectors to some hundred nm. The rich intermittent plastic activity admitted by the deforming crystals has now been readily demonstrated with micro-crystal stress-strain experiments of fcc [167-169], bcc [170, 171], hcp [172], and bcp [173] lattices, whereas historically it has been mainly limited to hcp lattices that strongly restricts deformation to the basal plane. Before turning our attention to the statistical insights of intermittent plastic flow in Section 4.2, it is noted that subsequent measurements on discrete deformation of Zn-crystals by Becker and Orowan [31] allowed to assess the time-scales associated with the observed displacement jumps. These are ca. 0.1-0.2 s, giving a slip velocity of the order of some μm/s, which is remarkably similar to the corresponding slip velocity obtained from velocity-profile analysis of deforming Au-microcrystals [166, 174]. Slip cinematography studies on Al by Becker and Haasen give similar values [175], whereas Anderson and Brown reported somewhat larger values of the order of tens to hundreds of μm/s using resistivity measurements on Zn crystals [176]. This very limited data therefore suggests that the underlying collective dislocation activity mediating slip in micro- and macro-crystals leads to a similar velocity range – even though the flow stress level is markedly different. Sampling the slip velocity across nano- and micro-crystals, which are known to exhibit a power-law size-dependent strength of type $\sigma \sim d^{-n}$, where $d$ is the crystal size and $n$ between 0.5-1 for fcc [41, 177], revealed no size and therefore stress dependence on the slip velocity [174]. Such a result can only be rationalized if the underlying dislocation dynamics is dictated by the internal stress components, rather than the applied stress. Bulk-scale 2D dislocation dynamics simulations (DD) gave insights into this problem, suggesting that it is only in the micro-plastic regime of a bulk stress-strain curve where a strong scaling between the average dislocation velocity and the applied stress is lacking [174]. In fact, this observation is also compatible with the earlier discussed creep-rate reported on by Chalmers and displayed in Figure 2, where only small changes in creep rate, and therefore dislocation velocity, occurred during micro-plastic deformation.

Whilst the micro-compression technique has reinvigorated interest in discrete plasticity, nano-indentation had revealed this behavior decades earlier and may actually also give valuable contemporary insights into intermittent plasticity ranging from the plastic evolution of a pristine crystal to that of a well-developed dislocation structure.

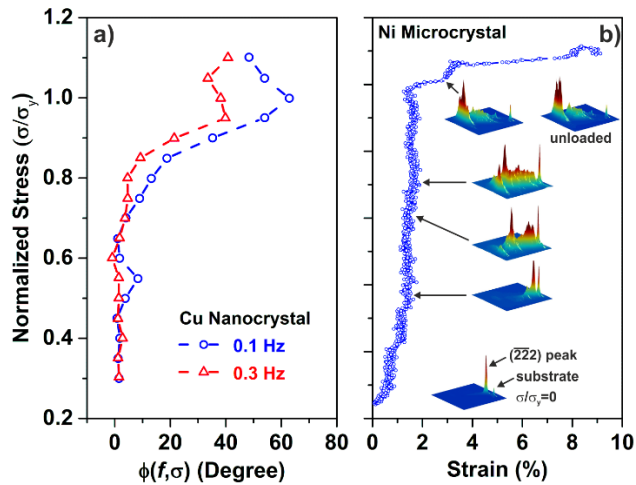

Figure 12: a) Normalized stress as a function of phase for two different frequencies obtained during deformation of a Cu nanocrystal [178]. b) Evolution of a $\overline{2}2\overline{2}$ Laue diffraction spot during deformation of a Ni microcrystal, indicating plastic deformation with the nominally elastic loading regime. In both experiments, plastic dissipation is revealed that begins at ca. $0.7 \times \tau_Y$.

Another noteworthy recent development is the combination of quasi-static and dynamic mechanical excitation during nano-mechanial testing. Directly linked to the briefly introduced internal friction measurements in Section 2.3, is the probing of dislocation activity with dynamical excitations in the nominally elastic regime of a small-scale deformation curve similar to the one shown in Figure 11. Ni et al. [178] reported for Cu nano-crystals that the cumulative oscillatory response prior to the apparent yield (break-away stress) indicates a dissipation signature that emerges at ca. $0.7 \times \tau_Y$ and peaks prior to the onset of extensive plastic flow (Figure 12a). This finding is strongly compatible with the earlier observation on a micron-sized Ni crystal obtained during in-situ Laue diffraction that a substantial dislocation structure may evolve in the early loading, which appears as an elastic response [168] (Figure 12b). At ca. $0.65 \times \tau_Y$, dislocation substructures formed, as was revealed from the angular spread and complex diffraction-peak intensity-landscape. Based on these findings, one may be tempted to break-down the "pre-yield" regime of the micro-crystal curve in Figure 11 into similar stages as done in Figure 1, and to identify a micro-plastic regime of the micro-deformation curve. If such a distinction can be shown to be sound, it naturally leads to the question as to how such a micro-plastic regime would be different from the bulk case.

*4.2 Scale-free slip behavior*

In very general terms, the aforementioned infrequent (but large) AE-bursts indicate non-Gaussian tails to the distribution of AE-signals, and therefore a partial breakdown of the assumption of a statistically meaningful average local environment. Later in Section 4.3, the same will be seen to apply to the slip-sizes extracted from small-scale intermittent flow curves, such as the one shown in Figure 11. It was the work of Miguel et al. [179] that first demonstrated non-Gaussian statistics for creeping ice undergoing basal slip in the form of a power-law distribution with a non-integer exponent. Such behavior was also found in metallic hcp materials in which a single slip system dominates [179] (it is worth noting that the early works reporting intermittent flow used Zinc crystals [21, 30-32, 176], in which a single slip plane governs the plastic response). For the found exponent, the pure power-law distribution entailed a divergent mean, indicating a lack of a well-defined energy scale. This was also reflected in the distribution of plastic event magnitudes, whose exponent is related to the AE energy burst spectrum. In the same work, this latter result was also seen in two dimensional DD simulations. Both the AE-pulse energy distribution and the energy-bursts distribution obtained from the collective velocity of all dislocations participating in an event are reproduced in Figure 13 for different creep stresses. The distributions are insensitive to the stress level, which also indicates that the underlying dislocation velocity, estimated from the AE-pulse via $E_{AE} = V^2$, is also insensitive to the applied stress. Here $E_{AE}$ is the pulse energy, and $V$ the collective dislocation velocity. This finding is in very good agreement with slip velocities directly obtained from micro-crystal deformation [166, 174, 180].

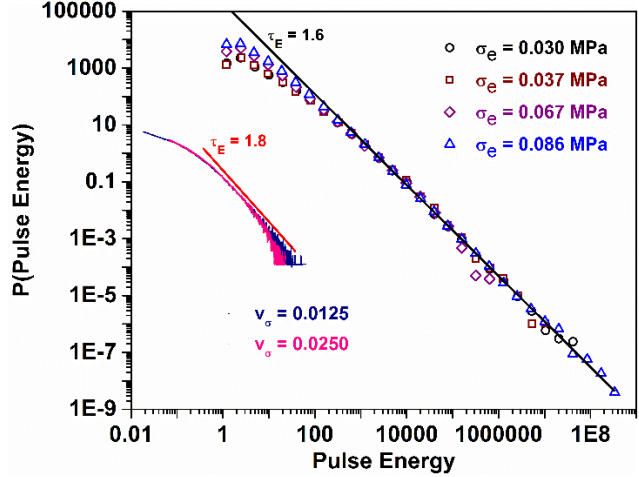

Figure 13: Statistical distribution of acoustic energy bursts recorded during creep experiments of ice single crystals showing a power-law scaling with an exponent of 1.6. 2D DD Simulations yield an exponent of 1.8. Reprinted/reproduced by permission from Macmillan Publishers Ltd: Nature [179], copyright (2001).

Strong power-law behavior can be a signature of criticality and scale-free behavior [181], and has in the context of plastic flow not only been reported for plasticity of crystalline materials, but also for shear-banding of metallic glasses [182], and other deforming disordered systems [183-185]. For a recent review see also Ref. [186]. In the present context such scale-free behavior manifests itself in a diverging dislocation-dislocation correlation length entailing collective activity of the underlying dislocation network occurs at all length scales. For the materials experiencing this type of deformation, the underlying plasticity mechanism represents a paradigm shift away from the unit-plastic mechanisms for which Gaussian correlations dominate.

In practice, a system exhibiting critical-like behavior admits a truncated power-law distribution of the form: $P(s) \sim (1/s^\tau) f[s/s_0]$, where (here) the plastic displacement magnitude, $s$, is explicitly considered. The scaling function $f[\cdot]$ controls how the power-law distribution is truncated at values of $s \sim s_0$ and is often the analytical function $\exp[-x^2]$. It is important to bear in mind that the truncation of the power-law naturally includes a lower and an upper limit; the former one given by either the experimental resolution or the finite size of a unit plasticity mechanism (which is ultimately related to atomic structure), and the latter one given by either an internal (static or dynamic) length scale, or a length scale associated with system size, as in the case of small crystals, where the slip length cannot exceed the physical dimensions of the material volume.

It is found that a broad range of systems, ranging from magnetism to earthquakes [187], display the above form for a variable which is related to some form of energy release. This motivates the terminology of "universality". In fact, those systems which have the same numerical value for the exponent and scaling function form, are said to belong to the same universality class. Since simple (analytically tractable) models may belong to a similar universality class as that of real complex metallic systems, it becomes justified to use simple models for the study of intermittent plasticity. This is somewhat different from the conventional materials science approach to modelling, which often involves the principle that the more complex and material specific the model or simulation is, the more realistic and applicable it becomes.

Within the above framework, two questions become relevant: 1) What are the value of the exponents?, and 2) What controls the truncation length-scale? Both questions relate to which universality class a material belongs to. A very relevant example is the case of the mean-field universality class. This simplified picture assumes all degrees of freedom - no matter their spatial separation - have the same interaction energy (an idea that is loosely based on the long range nature of the dislocation-dislocation interaction [188]). Because of this, an arbitrarily chosen dislocation segment will interact with an un-weighted average of all local dislocation environments admitted by the system, thus motivating the terminology of the "mean-field". The mean field assumption has been used in cellular automata models of plasticity [189], interface depinning in a field of quenched disorder [188], and zero-dimensional non-linear visco-plastic models [154, 190]. For the exponent associated with the plastic event magnitudes, this universality class gives a value equal to precisely $\tau = 3/2$, which is very close to the value found for creeping ice ([179], Figure 13) and often obtained in DD simulations [179, 191-194].

The question to which universality class plasticity (prior to yield) belongs to has obvious relevance to the physics of micro-plasticity discussed in this article. The work of Dahmen [190] and Mehta [195] employs a zero-dimensional "tuned criticality" mean-field depinning model of plasticity in which the system is only at criticality when the external stress is equal to the depinning stress. At this external stress value the dislocation-dislocation correlation length diverges (and is thus truncated only by system size). On the other hand, for external stresses lower than the depinning stress – the regime of micro-plasticity – the dislocation-dislocation correlation length is finite and controlled by a power-law with respect to the difference between the applied stress and the yield stress. In this scenario, the dislocation structure prior to loading is far from criticality so initial (micro-)plasticity can be Gaussian with a well-defined length-scale. As the loading proceeds, the dislocation structure reorganizes itself and the dislocation-dislocation correlation length grows to eventually diverge at a critical stress. The zero-dimensional model of Dahmen and co-workers does have some experimental support in Ref. [196], which investigated the size-dependent statistics of discrete plasticity in nano/micron-pillar deformation experiments. A simplified dislocation dynamics model in which the immobile dislocation content is treated statically via an imposed internal stress and length scale, also exhibits this "tuned" criticality [193, 197].

The above picture has however recently been questioned [198], in which both cellular automata, and two and three dimensional dislocation simulations show a non-mean-field set of exponents which may be associated more with the phenomenon of unjamming [199, 200]. In this work, $s_0$ scales as a power-law (exponent $\beta$) with respect to the system size and therefore will diverge for a bulk system implying that the system is in a state of criticality for the entire micro-plastic regime and no length and energy scale (apart from that associated with sample size) can be assigned to the discrete plasticity. Such criticality is referred to as self-organized criticality [201]. $\beta$ is however below 1/2, and the prefactor of the scaling is an exponential function of the applied stress $\sigma_{ex}$: $\exp[\sigma_{ex}/\sigma_0]$. Thus, whilst there is criticality at all stress levels, $s_0$ can be small for low stresses and finite (but large) sample sizes.

These differing conclusions indicate that to which universality class (or classes) plasticity (and therefore micro-plasticity) in metals belongs to, remains an unresolved question.

As already stated, many of the materials which exhibit scale-free behavior have a dominant slip plane. For the case of dislocation networks in which many slip systems are active – where there is strong network entanglement such as in work hardened metals – discrete activity appears to be both diminished and less power-law like [202]. Here, $s_0$ is set by an internal length-scale characterizing the network. This may also arise from introduced internal length-scales such as impurity concentrations in alloy systems [203]. The recent work in Ref. [202], suggests for such metals that this correlation length is sufficiently small for Gaussian correlations to dominate. One important consequence of critical behavior is that the response of the system (here, in terms of plastic displacement) is not deterministically connected to the stimulus (here, the application of load). From an engineering perspective this is clearly not desirable. Fortunately, such a system might be said to be the exception rather than the rule, since internal length scales generally emerge within dislocation structures resulting in finite dislocation-dislocation correlation lengths and thus material dependent values of $s_0$ that might relate to dislocation network, cell, and grain boundary length-scales, or dislocation segment length-scales that ultimately lead to the traditional concepts based on dislocation bowing and break-away stress being the relevant unit scale plastic processes. However, our understanding on what internal structural length-scales limit the observable maximum event sizes is rather limited.

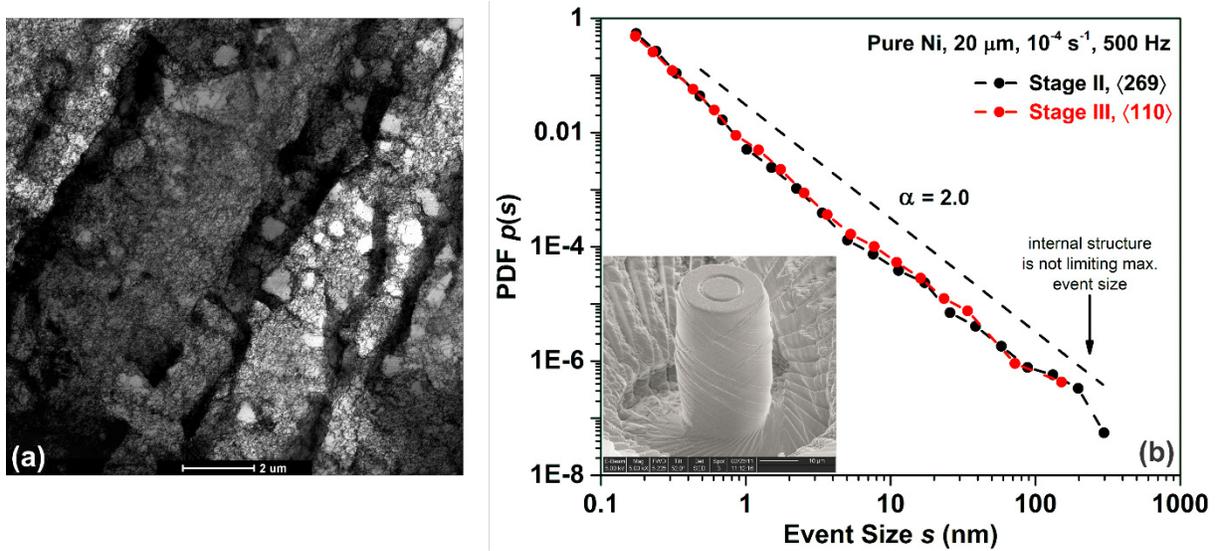

Figure 14: a) Dislocation structure as obtained after pre-deformation into stage III hardening (23% strain) of a pure Ni bulk crystal. Microcrystals with a diameter of 20 μm were prepared and compressed [204]. The resulting event-size distribution is shown in b), revealing that despite the pronounced pre-existing dislocation structure, the distribution remains unaltered in both slope and largest observed event size. The inset in b) displays the deformed microcrystal [204]. Unpublished data, kindly provided by D. Dimiduk and co-workers [205].

This aspect has been highlighted by the recent work of Dimiduk and co-workers [205] who investigated the relationship between internal structural length scales and truncation effects for Ni micro-crystals of two different micro-structural states. Those were pure ⟨269⟩-oriented micro-crystals deformed through stage II hardening [206] with an approximate dislocation density of $\sim 1 \times 10^{12}$ m$^{-2}$ [167], and ⟨110⟩-oriented micro-crystals prepared from a bulk crystal that was pre-strained by 23% into stage III hardening [204]. The pre-straining of the bulk material resulted in a dislocation density of $\sim 4 \times 10^{14}$ m$^{-2}$, which is about 6 orders of magnitude higher than in the as-received un-deformed bulk crystal. The resulting dislocation structure of this pre-deformed material is shown in Fig. 14a, clearly indicating dislocation wall formation, dense dislocation patterns, and some cellular structures. It is remarkable to see that the resulting slip-size distributions from the two microcrystals, with very different initial dislocation structures, do not have a different size cut-off for the event size, as shown in Fig. 14b. These findings clearly underline that the development of dislocation structures with an increasingly smaller characteristic internal length scale need not limit the maximum event size. In other words, the mean free path of the dislocations can be larger than an identified structural length scale.

*4.3 Is small scale-plasticity similar to bulk micro-plasticity?*

With the advent of very sensitive displacement sensors and load transducers it has become more routine to measure sample displacements of the order of nano- and sub-nano-meter and changes in force of the order of micron- and sub-micron-Newtons. This was discussed in Section 4.1, and a direct comparison between a macroscopic and microscopic deformation response was made in Figure 11.

Whilst the displacement resolution of such modern devices is exceptional, the effective strain resolution is low when compared to that of the ultra-high strain resolution of the torsion apparatus of Refs. [20, 21]. That such devices are able to resolve discrete plastic events is mainly due the small volume of the plastic zone. This may be seen via the relation (originating from the Eshelby inclusion construct [136]) relating the plastic strain of *bulk* plasticity associated with two different plastic volumes, $\varepsilon_{\text{micro}}^{\text{plastic}} V_{\text{micro}} \simeq V_{\text{bulk}} \varepsilon_{\text{bulk}}^{\text{plastic}}$. Taking the bulk values from that of (say) Tinder and co-workers $V_{\text{bulk}} \approx 1$ mm³ and $\varepsilon_{\text{micro}}^{\text{plastic}} \approx 10^{-8}$, a comparable plastic strain in a volume of $V_{\text{micro}} \approx 1$ μm³ will be $\varepsilon_{\text{micro}}^{\text{plastic}} \approx 10$. A simple usage of this scaling implies that plasticity observed in micron-deformation experiments may be cast within the framework of bulk plasticity. In turn, this viewpoint has the natural consequence that the discrete activity seen in such experiments has its origins in the micro-plastic regime prior to yield in a bulk system; a conclusion brought forward in Ref. [174] and also outlined in Section 4.1.

In the prominent work of Uchic and co-workers [40, 207], flat-punch indentation of focus-ion-beamed single crystal micro-pillars revealed, in addition to the popular extrinsic sample size effect briefly mentioned in Section 5.1, increasing discrete stochastic plastic activity as the pillar diameter reduced in size. In later work [206, 208-212], inspection of the statistics of the associated discrete plastic displacement magnitudes suggested scale-free dislocation avalanche activity with an exponent for the event size distribution similar to that of mean-field theory: 3/2. A study by Diminduk [213] on LiF microcrystals revealed significant deviations away from mean-field exponents and reported rather Gaussian-like behavior for smaller events, whereas the larger events had power-law forms over certain regimes. All these studies on event sizes did not consider the statistics as a function of the stress (relative to some yield stress value), however later work on submicron crystals did just this [196], suggesting the discrete plasticity could be described by the tuned-criticality of a mean-field depinning transition [190, 195].

Not only the statistics of event sizes measured by micro-compression of small-scale crystals, but also the corresponding time-scales have been subject of experimental investigations [166, 174]. Together these quantities reveal a slip-velocity distribution that covers two orders of magnitude with a distinct flat shoulder at low velocities, and a high-velocity tail with a cubic decay, as shown in Figure 15a. The same distribution form was reported for dislocation avalanche velocities obtained from 3D DD of sub-micron crystals [192]. Furthermore, the cubically decaying tail is compatible with mean-field predictions (dashed line in Figure 15a) [214]. As already noted in Section 4.1, the surprising observation of the event velocities recorded for deforming micro-crystals is that their values are remarkably compatible with the (very limited) data obtained from intermittent bulk deformation [31, 175, 176]. An additional finding that supports the validity of the mean-field universality class is shown in Figure 15b, where the time-resolved velocity profiles of slip events within different slip-size bins from deforming Au-microcrystals [215] have been collapsed onto the predicted scaling function for slip avalanches provided by Ref. [214]. Clearly, large fluctuations are present, but the averaged profiles from the fcc-lattice are well described by the mean-field prediction. Recent insights into average velocity profiles from other lattice systems suggest the need for different average shape functions in order to capture significantly different velocity-relaxation profiles than shown here for fcc [214]. In fact, lattice-system dependent avalanche velocity relaxation indicates that (at least) different universality classes may be at play for the dynamics of avalanches, even though their slip-size distributions seem to be well described with the same power-law exponent.

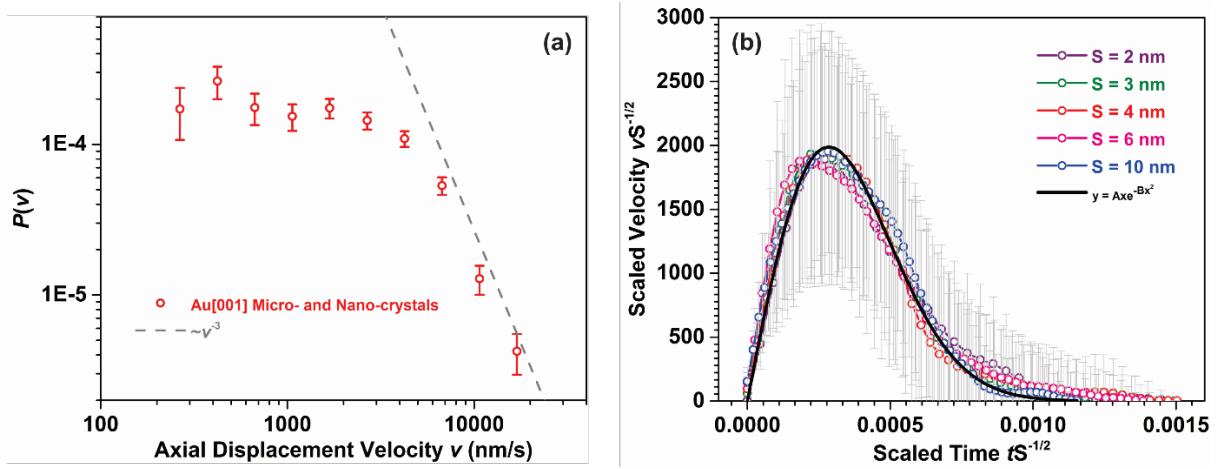

Figure 15: a) Distribution of axial displacement velocity, evidencing a shoulder at low velocities and an approximate cubic decay at higher velocities [166, 174]. The dashed line is the predicted scaling after Ref. [214]. b) Rescaling velocity profiles for different slip-size bins that collapse very well onto the predicted mean-field function $Axe^{-Bx^2}$ [214, 215], where A and B are fitted to 1.15×10$^7$ and 6.15×10$^6$, respectively. The data in both graphs was obtained from micro-compression on Au micro-crystals.

Whilst not explicitly discussed in the above experimental papers, the observation of power-law like slip-size distributions and the usage of criticality to rationalize them necessarily implies a certain type of bulk plasticity, and therefore a certain type of bulk micro-plasticity, is at play. This implication is quite intriguing, because crystals with diameters as small as 75 nm have been reported to exhibit scale-free power-law slip size distributions (Figure 16) [196]. Given that the size-effect in strength of these small crystals is often explained through a depletion of (mobile) dislocations, observing power-law scaling is not at all straight forward, unless the crystals still contain some relevant bulk-like dislocation structure. That bulk-like micro-plasticity is at play is of course masked by the very large strain jumps seen in the micron sized samples. Under the assumption of critical behavior for the entire deformation, the finite volume of the sample enters as the relevant truncation length-scale in the scaling function in $P(s) \sim (1/s^\tau) f[s/s_0]$. For the case of tuned criticality, when the stress is sufficiently low (far from an assumed critical depinning stress) the dislocation-dislocation correlation length is less than that of the extrinsic length scale. In this regime sample size should not enter into the scaling function. However, at high enough stresses (corresponding to large enough correlation lengths) or small enough system sizes, the extrinsic length-scale will be the cause of truncation. This regime of small enough system sizes was considered to be of the order of a hundred nanometers [196], as suggested by Figure 16.

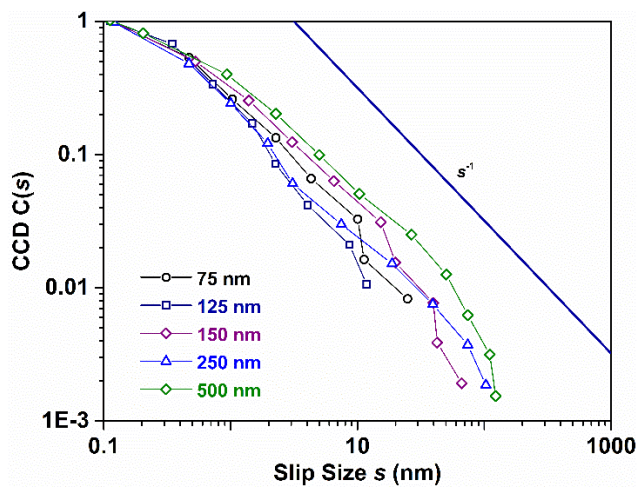

Figure 16: Complementary Cumulative Distribution (CCD) of slip sizes obtained during micro-compression of Au crystals. Reprinted/reproduced with permission from Ref. [196]. Copyright (2012) by the American Physical Society.

The recent work of Weiss *et al*. [202] finds, via AE-experiments, that for traditional bulk metallic materials with well populated slip-systems, Gaussian fluctuations dominate the plastic strain statistics of bulk plasticity. The observation of scale-free statistics in micron-sized metallic samples is rationalized by asserting that because of the presence of the surface, dislocation number and entanglements are reduced, promoting single slip behavior and therefore a dominance of power-law statistics whether it might be critical or tuned-criticality. From this conclusion, it can be inferred that reducing sample size replaces a bulk plasticity dominated by Gaussian fluctuations with a particular bulk plasticity now dominated by power-law statistics. This offers a rather general explanation of plasticity in reduced volumes without an explicit reference to a change of mechanism, as has often be proposed in the body of work focusing on extrinsic sample size effects of microcrystals [38, 177, 216], which will be discussed further in the proceeding section.

*4.4 Statistics of critical stresses*

Sections 4.1-4.3 outline work concerned primarily with the stochastic aspects of the plastic strain magnitude, energy release and velocity magnitude of individual plastic events. Less work has been done on the stresses at which such events occur. Indeed, the observation of the stochastic nature of intermittent plasticity motivates an entirely probabilistic description of plasticity, in which the statistics of all observables is ultimately based on an ensemble of starting micro-structures. This is ideally suited for the modelling of intermittent micro-plasticity, since inherent to the idea of micro-plasticity is that over a finite number of plastic events, there is little change in the parameters that define the microstructure. This viewpoint is also supported by the classical experiments discussed in Section 2, and some evidence gained from in-situ Laue diffraction [168] as well as TEM [217] work on micro-crystals suggests that once plastic flow is recorded, the dislocation structure (in terms of dislocation boundaries, rotational gradients, and density) does not noticeably evolve, even though there obviously is a strong net flux of dislocations that admit plastic deformation. With each plastic event, the microstructure does change, however the parameters defining the statistics of microstructural realizations (the master distribution) are expected to change little - in other words hardening is negligible. In practice the ensemble underlying the statistics would constitute micro-structural realizations arising from a sample preparation protocol which would impose constraints such as a fixed dislocation density, or an internal length-scale associated with a well-defined dislocation structure.

Within the above framework, a master distribution of critical stresses is one way to characterize the ensemble of microstructural realizations over which the statistics is made. The construct of a master distribution of critical stresses, $P(\sigma)$, has been used in past works [41, 197, 218-226]. For example, Ngan and co-workers [218-222] have developed a general probabilistic description for the statistics of the *m*th plastic event occurring at time *t*. Under a constant stress rate loading geometry ($\sigma(t) = \sigma = \dot{\sigma}t$), their work considers the probability that a member of the ensemble has undergone *m* plastic events at stress $\sigma$, $p_m(\sigma)$, giving the average number of discrete plastic events as $N(\sigma) = \sum_{m=0}^{\infty} m p_m(\sigma)$. The probability, $p_m(\sigma)$, can be written as, $p_m(\sigma) = F_m(\sigma) - F_{m-1}(\sigma)$ [220], where $F_m(\sigma)$ is the survival probability that the sample has not undergone the *m*th plastic event at stress $\sigma$ (with $F_0(\sigma) = 0$ and $F_\infty(\sigma) = 1$), giving $N(\sigma) = -\sum_{m=1}^{\infty} F_{\bar{m}}(\sigma)$. In Ref. [222], $\dot{N}(\sigma)$ is assumed to be proportional to $VP_<(\sigma)$, the average total number of plastic events that have occurred up to the stress level $\sigma$. Here $V$ is the volume of the sample and $P_<(\sigma) = \int_0^\sigma d\bar{\sigma} \, P(\bar{\sigma})$ is the cumulative probability density distribution per unit volume of a critical stress for a plastic event to occur – where $P(\sigma)$ is the above introduced master distribution. Such a distribution was assumed to originate via $\sigma \approx \mu b/L$ from a distribution of Taylor segment lengths [222].

Whith $VP_<(\sigma) \sim 1$ the stress scale at which the first plastic event occurs is defined. Indeed, for a power law $P(\sigma) \sim \sigma^\alpha$, $VP_<(\sigma_1) = 1$ gives a characteristic stress of $\sigma_1 \sim \left(\frac{1}{V}\right)^{\frac{1}{\alpha+1}}$. This is analogous to a central result of the extreme-value- and order-statistics approaches in which a master critical stress distribution is sampled $M$ times, with the characteristic lowest critical stress scale, $\sigma_1$, being given by $MP_<(\sigma_1) = 1$. For sufficiently large $M$ (and therefore sufficiently small $\sigma$ for which $P(\sigma) \sim \sigma^\alpha$), the average critical stress associated with the first plastic event is given by $\langle \sigma_1 \rangle = \sigma_1 \Gamma[2 + \alpha]$, with $\sigma_1 \sim \left(\frac{1}{M}\right)^{\frac{1}{\alpha+1}}$, and the fluctuations around this value are described by a Weibull distribution with scale parameter equal to $\sigma_1$ and shape parameter equal to $\frac{1}{\alpha+1}$. By assuming $M = \rho V$ where $\rho$ is a density of available plastic events, a very fundamental size effect in stress again emerges. This formalism has been shown to describe very well the statistics of the first pop-in event in nano-indentation data as a function of the radius of the spherical indenter (see Figure 17), as well as that of the micro-plastic regime of a dislocation dynamics simulations [197, 226]. For subsequent plastic events, $\sigma_1 \sim \left(\frac{1}{V}\right)^{\frac{1}{\alpha+1}}$ has been generalized to $\sigma_i \sim \left(\frac{i}{V}\right)^{\frac{1}{\alpha+1}}$, where $i$ indexes the *i*th discrete plastic event [41]. Such an expression is found to work well for both a stochastic continuum plasticity and discrete dislocation dynamics model [227].

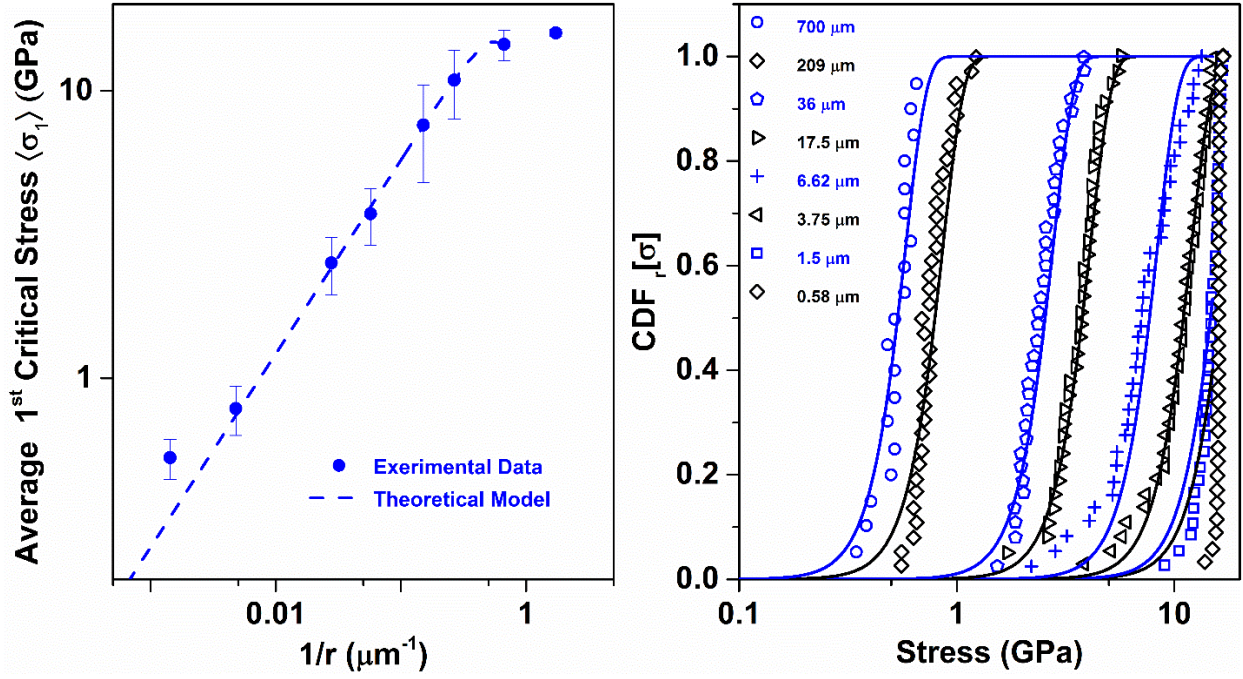

Figure 17: a) Experimentally measured average stress of the first pop-in during nanoindentation on Mo single crystals as a function of indenter radius [156]. The line is a fit of a theoretical model proposed in Ref. [226]. b) Cumulative distribution function (CDF) of the same nanoindentation data as in a) with the predicted Weibull CDF derived theoretically [226]. Reprinted/reproduced with permission from Ref. [226] with the permission of AIP Publishing.

There exist a number of works which have exploited the extreme-value-statistics size effect in stress to rationalize the small-is-stronger effect seen in micro-crystals [41, 222, 224], giving a size effect coefficient depending in part on alpha. In the work of Refs. [41, 224], the size effect in stress is combined with a size effect in strain, to give a size-effect coefficient equal to $n = (\tau + 1)/(\alpha + 1)$ when the system is in a state of criticality, and no dependence on size when the system has a dislocation-dislocation correlation length less than external sample dimensions. This approach gives a rather simple understanding of the extrinsic size effect (smaller-is-stronger) [41] that requires no specific mechanism and therefore could equally well be applied to the grain-size versus strength effect epitomized by the Hall-Petch relation [224] (Figure 18). Indeed, its only requirement (in addition to an extreme-value-statistics approach for the critical stresses) is that the plastic strain events are statistically independent of each other and are dominated by finite size effects due to criticality - an assumption which appears to be valid for micron-sized metallic samples [202, 206, 208], but requires modification in order to capture the weaker anomalous system size dependence of two dimensional dislocation dynamics systems [227]. As discussed in Section 4.3, the recent work of Weiss and co-workers [202] suggest that in this crystal volume regime, the presence of a surface (or interface?) might induce (through dislocation reorganization) critical behavior and thus, according to Refs. [41, 224], a size effect. A relevant mechanistic manifestation of this might be the emergence of single-arm dislocation sources that have been seen in both simulation and experiment [228]. The work of Refs. [41, 224] also suggests a richer scaling in terms of sample shape since the extreme-value-statistics size effect in stress will depend on volume and the size effect in strain will depend on accessible slip area, resulting in a scaling that depends on aspect ratio when at a fixed pillar diameter [225].

It is emphasized, that such scalings with system size only describe a leading order algebraic volume dependence. In Refs. [41, 224, 225], it has been argued that when considering the size-effect in terms of a log-log plot of stress versus an extrinsic or intrinsic length scale (see Figure 18), only logarithmic (and therefore power-law) accuracy is needed to capture the observed trends. From this perspective, it is the analytical prefactors that can depend on volume, plastic strain, strain rate, and temperature in a non-universal material-specific way, leading to the strong scatter between materials typically seen in figures such as Figure 18.

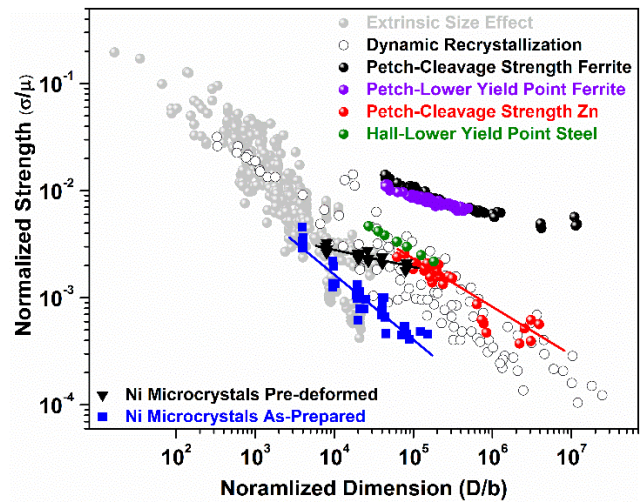

Figure 18: Log-log plot of strength and dimension (internal or external length scale) for a wide range of small-scale crystals and grain-size data, including the original data from Hall and Petch. The normalization of the axes was done as proposed by Derby [229]. Reprinted/reproduced from Ref. [224] with permission from Elsevier.

What is the origin of $\alpha$? Many strongly interacting systems follow a power-law distribution in a quantity that characterizes the onset of a driven instability. In the context of marginal stability, such power-law distributions are referred to as psuedo-gap functions [230]. Importantly, if the system is in a state of criticality prior to and during initial loading, the exponent $\alpha$ may have a saturated value that is related to the universality class of the criticality. If the system is not in an initial state of criticality, and only through loading does it organize to a (tuned) critical state, then in the micro-plastic regime, $\alpha$ will be non-universal and very much dependent on material specifics and its preparation protocol, as well as the specific plastic evolution towards yield. It is presently unknown what would be the saturated value of $\alpha$ for a dislocation network. Recent analysis has shown that its value (whether saturated or not) can be approximately 0.3-0.4. This is obtained from first pop-in experimental nano-indentation data [226], a simplified one-dimensional dislocation dynamics model [197, 226], and a two-dimensional dislocation dynamics system [227]. On the other hand, a value of approximately three is obtained when using $n=(\tau+1)/(\alpha+1)$ to rationalize the size effect [41, 224], from small-scale experiments on nanocrystalline Ni [231] and single crystals [232], or through detailed analysis of TEM cross-sections of (albeit 2D) dislocation structures [222].

## 5 Conclusions and future research directions

This review article spans many decades of focused research on fundamental aspects of early plastic flow, here categorized as micro-plasticity. The long history of this field allows us to barely scratch the surface of the many contributions, and we have here selected a few which motivate the connection between classical micro-plasticity and modern developments in intermittent, scale-free flow of macro- and micro-sized crystals, and metallic glasses. The aim of the foregoing sections was to discuss the development from classical deformation theory, describing flow with unit mechanisms and statistically meaningful average quantities, to the current focus on scale-free behavior that does not exhibit well-defined length and stress scales. This paradigm shift is mainly driven by observations of plastic flow obtained from very small deformations of bulk crystals, or from modern small-scale mechanical testing, which i) motivated our focus onto micro-plasticity of bulk crystals, and ii) suggests a fundamental link between the stress-strain response of miniaturized specimen and bulk micro-plasticity. To broaden the scope of this article, we have included micro-plasticity of metallic glasses, for which it becomes increasingly clear that there is a large amount of structural plastic processes occurring at virtually no stress even though these materials have a very repeatable bulk elastic-to-plastic transition. These new insights suggest that it is the micro-plastic regime in metallic glasses that can be used to homogeneously tune the structural state and enthalpy of the system, whereas reported microstructural changes due to static loading below macroscopic yield seem minimal in crystals. This fundamental difference between micro-plasticity in crystals and metallic glasses opens up, for each material class, promising avenues where micro-plasticity can be used to improve our macroscopic materials understanding.

We, however, believe that the most exciting novel research directions lie in the field of atomistic and microscopic structural behavior during (thermally activated) micro-plasticity. Coined in 1932 by Chalmers, the precise micro-plastic behavior of both crystals and metallic glasses remain largely unknown, even though they are of paramount importance for the true definition of the elastic limit, how a defect structure develops towards yield, how metallic glasses rejuvenate or relax, how mesoscopic shear defects (shear bands) form in metallic glasses, or for how statistical physics can be adequately used to develop coarse-grained flow models on the basis of an approximate scale-free plasticity. It will take time to unravel the details of micro-plasticity in crystalline and amorphous metals, and (on the basis of contemporary work) we conclude with proposing a few questions that may help us on our way:

- Prior to encoding the detailed microstructural evolution during micro-plasticity, an effort needs to be made to properly define the boundaries of micro-plastic flow. In particular, the transition to macroscopic flow has not received much experimental attention. With modern sensitive mechanical testing equipment and state-of-the-art synchrotron facilities (for example x-ray topography), a much more refined view on how a defect population develops towards (an engineering) yield should be possible and may provide a definition of yield based on more fundamental parameters than the 0.2% offset strain. This has not only a value for stress-strain behavior in general, but links directly to modern statistical approaches to quantify plasticity. In particular, quantifying the microstructural evolution towards yield will clarify if the material gradually develops towards criticality, or if a representative initial defect structure (dislocation network) is so from the beginning.
- The previous point puts focus on the statistical assessment of not only the discrete plastic strain increments, but also the critical stresses at which they occur. Strong focus has hereto been on the statistics of the first event, which was often linked to heterogeneous dislocation nucleation or shear-band formation, but how does the statistical signature of critical stresses evolve for higher order events? In particular what is the value of the exponent of the asymptotic critical stress distribution? How does it evolve with each event, as the yield transition is approached? What will this teach us about the evolution of the defect structure? Experimentally, this could be achieved by a careful effort in characterizing pop-ins in nano-indentation experiments beyond the first event, or through the deformation of micro-crystals which exhibit intermittent flow but (because of their large enough size) do not yet display a size dependence in strength. This latter direction might be best done in tension using for example micro-wires.

- In Sections 4.3 and 4.4 we raised the question if bulk micro-plasticity is the same – or similar – to plasticity at small scales. A satisfactory answer to this question remains elusive, and its resolution may incorporate some provocative aspects. Classical work may provide the appropriate starting point to resolve this question, such as the finding that little microstructural change occurs during micro-plasticity. This result could be rigorously tested through a systematic study of dislocation density, collective velocities and more generally the spatio-temporal dynamics of intermittent plasticity of crystals that show the typical intermittent flow, but that do not yet display any size-dependence in strength.

- The phenomenon of criticality has been very useful in understanding the scale-free signatures of plastic flow. One unresolved question is how such power-law scaling is truncated. What are the relevant length scales that determine the truncation length scales? Are they dynamic or static? This question can also be related to the universality class of the criticality. Some evidence speaks for static length scales such as grain size, other internal (structural) length scales, or finite crystal sizes, but given the result of Figure 14 we have to critically ask ourselves if it may not be generally related to traditional structural length-scales? For example, will a collective dislocation rearrangement (avalanche) in a polycrystal be limited to its confining grain, or can a cascade of avalanches in adjacent grains emerge? High-speed x-ray diffraction or spatially resolved tracking of surface step formation during straining could be avenues to seek answers to these questions.

- Intermittent flow has been primarily described by the statistics of acoustic emission data or slip-size magnitudes. Some early work in this field [179, 233] motivated new plasticity models that incorporate the notion of scale-free statistics rather than homogenization on the basis of mean-properties. We believe that by viewing an avalanche as a dynamical object [166, 174, 180, 215], which can be experimentally characterized and quantified, will provide a useful bridge between statistical and mechanistic descriptions of plasticity and thus help in the development of new predictive flow models. Indeed, understanding an avalanche in terms of crystallography, micro-structure and temperature would parallel the original approach of Orowan and Becker (Section 1.2), which aimed at understanding the mechanisms and origins behind crystallographic slip-velocities.

- As discussed by Zaiser [154, 210], it is a remarkable finding that the statistical signature of slip in fcc and bcc micro- and nano-crystals suggests the same universality of strain-burst behavior even though the flow-stress controlling mechanisms are thought to be very different. This emphasizes an important implication of the modern scale-free view of plasticity, that is, a wide class of materials can belong to the same universality class. This seems to go head-to-head with the long and rich history of the material-specific discipline of materials science. Clearly, if all scaling exponents are found to be the same, it is hard to envision the development of plasticity models for differently flowing materials. A quantitative framework that brings together these conflicting implications would be an important development. Why is a material dominated by Gaussian statistics and another by scale-free statistics? Can a mechanistic description span both regimes of behavior? Possible ways forward are entailed in the previous two points, but also a community effort (currently being done) to determine the universality class or classes of the materials we all study is necessary.

- This review focuses on micro-plasticity and its connection to developments in intermittent plasticity. Even though scale-free intermittent flow has experimentally been revealed on the basis of small strain experiments of bulk crystals, it is expected (and has also been numerously shown) that intermittency can prevail throughout macroscopic flow as a general property of plasticity. This naturally leads to the question how failure initiation can be understood and described by dislocation avalanches? Is it possible to predict failure by knowing the time series of pre-cursor intermittent plastic events? Does there exist distinguishable precursor activity? In other words, is there a particular part of the distributions as shown in Figures 13-16 that will drive, for example, crack formation?

- Properly characterizing micro-plasticity will not only need statistical protocols, but also an effort to include thermal activation. This applies to both crystals and metallic glasses. What remains unclear is how to rationalize thermal activation parameters that are fundamentally averaged quantities, with scale-free descriptions of plasticity. This will not only require the inclusion of temperature (or strain rates) into non-equilibrium physics models for plasticity, but to also make a detailed experimental effort to probe temperature (or rate) dependencies of micro- and small-scale plasticity.

- In the special case of metallic glasses, the discussed data suggests that the micro-plastic regime offers a promising means to tune the structure in terms of enthalpy and therefore volume per atom. How much can the structure be modified, or shifted up along the enthalpy curve? If enthalpy storage (that is, overall macroscopic rejuvenation) of a similar level as seen in inhomogeneously deformed glasses is within reach, micro-plastic flow can be the key to explore the maximum possible enthalpy (and therefore structural) range accessible to the glassy state. Micro-plastic protocols may be combined with thermal strains induced by cooling to achieve, as yet, undiscovered thermomechanically tailored glasses. Such endeavors will also require the development of atomistic simulation protocols that are able to capture the collective rearrangements needed at low temperatures – again emphasizing the need for thermal activation [138].

With the above suggestions for future research directions, and the therefrom emerging connection between classical micro-plasticity and modern topics in plastic flow, we conclude by noting that the two somewhat contradicting views on plasticity need to be brought into accordance. Both have their regimes of validity, and both allow to understand (and sometimes predict) specific aspects of the plastic response of a material. However, a physics-based, rather than phenomenologically-based, predictive plasticity model for the stress-stain response of a given material that includes both aspects, has, despite more than 100 years of research, not yet been achieved. In this context, the marriage of the here discussed views should be seen as an opportunity rather than a conflicting understanding of the same evolutionary problem.

# 6 Acknowledgements

We thank the editorial board of Acta Materialia for being given the opportunity to write this article. Karin Dahmen, Konrad Samwer, Dennis Dimiduk, and Gregory Sparks are gratefully acknowledged for providing data and illustrations used in this review. Parts of the data shown here was obtained in the Frederick Seitz Materials Research Laboratory Central Research Facilities, University of Illinois at Urbana-Champaign. R.M. is thankful for startup funding from the Department of Materials Science and Engineering at UIUC, and for funding from the National Science Foundation (NSF DMR MMN Early Career Award, grant No. 1654065). The authors thank A. Beaudoin, K. Dahmen, D. Dimiduk, C. Woodward, P. Ispanovity, W. Curtin, and D. Rodney for insightful discussions.